# Ultraviolet optical horn antennas for label-free detection of single proteins

Aleksandr Barulin,[1] Prithu Roy,[1] Jean-Benoît Claude,[1] Jérôme Wenger [1,*]

[1] *Aix Marseille Univ, CNRS, Centrale Marseille, Institut Fresnel, 13013 Marseille, France*

\* *Corresponding author:* jerome.wenger@fresnel.fr

**Abstract:**

Single-molecule fluorescence techniques have revolutionized our ability to study proteins. However, the presence of a fluorescent label can alter the protein structure and/or modify its reaction with other species. To avoid the need for a fluorescent label, the intrinsic autofluorescence of proteins in the ultraviolet offers the benefits of fluorescence techniques without introducing the labelling drawbacks. Unfortunately, the low autofluorescence brightness of proteins has greatly challenged single molecule detection so far. Here we introduce optical horn antennas, a dedicated nanophotonic platform enabling the label-free detection of single proteins in the UV. This design combines fluorescence plasmonic enhancement, efficient collection up to 85° angle and background screening. We detect the UV autofluorescence from immobilized and diffusing single proteins, and monitor protein unfolding and dissociation upon denaturation. Optical horn antennas open up a unique and promising form of fluorescence spectroscopy to investigate single proteins in their native states in real time.

**Introduction**

One of the ultimate goals of molecular biology is to watch how single proteins work in their native state. While single molecule fluorescence techniques have achieved impressive results towards this goal,[1,2] the requirement for fluorescent markers can potentially lead to severe issues altering the protein structure or modifying its reaction with other species.[3–8] Therefore, label-free alternatives to detect single molecules are actively investigated.[9–13] The protein autofluorescence in the ultraviolet (UV) is an appealing route to rule out all the issues related to external fluorescence labelling.[14,15] More than 90% of all proteins contain some tryptophan or tyrosine aminoacid residues which are naturally fluorescent in the UV.[15] Being able to detect the UV autofluorescence from a single (label-free) protein would be a disruptive method offering many benefits of fluorescence techniques (signal-to-noise ratio, temporal dynamics, sensitivity…) without introducing the labelling drawbacks.

However, proteins are orders of magnitude dimmer as compared to conventional fluorescent dyes, so that single protein UV detection has remained a major challenge so far.[16–18] Hence, new nanotechnology tools need to be introduced to intensify the emission from single proteins. One of the main limiting issues is that close to a planar dielectric interface, a significant fraction of the light from a single dipole is emitted at large angles above 65°. This fundamental phenomenon is known as supercritical or forbidden light.[19,20] Microscopes operating in the visible spectral range use objectives of high numerical apertures of 1.4 or above to maximize the fluorescence collected from a single molecule. In the UV, however, the choice of microscope objectives is strongly restricted.[14] UV objectives have a numerical aperture typically below 0.8, which corresponds to a maximum collection angle of 33° into the quartz substrate of 1.48 refractive index. Collecting the forbidden UV light emitted at high angles is crucial to maximize the autofluorescence signal and unlock single label-free protein detection.

In analogy to radiofrequency antennas, optical antennas offer a way to control and intensify the emission of single quantum emitters.[21] Intense fluorescence enhancement factors have been achieved with strongly absorbing dyes in the visible,[22–27] but most optical antennas designs remain unsuitable for UV protein detection due to their narrowband spectral response,[28,29] challenging nanofabrication,[24,27] or requirement for solid-state integration.[30,31] Alternative designs must be developed to offer a highly efficient platform, reaching the needs of high photon count rates, microsecond time resolution, background-free operation and full compatibility with the UV detection of proteins.

Here, we introduce an optical horn antenna platform for label-free detection of single proteins in the UV with unprecedented resolutions and sensitivity. Our approach combines (i) a conical horn reflector for fluorescence collection at ultrahigh angles with (ii) a metal nanoaperture for fluorescence enhancement and background screening. To experimentally demonstrate the usefulness of our approach and its direct application to biochemical challenges, we detect the UV autofluorescence signal from immobilized and diffusing single proteins, and we monitor the unfolding and dissociation upon denaturation of a widely used protein. Optical horn antennas open up a promising form of fluorescence spectroscopy to investigate single proteins in their native states in real time. As additional advantage of our dedicated design, the improved brightness achieved with our optical horn antenna enables a 100-fold reduction of the experiment integration time as compared to the confocal reference. While the horn antennas are primarily developed here for UV protein detection, the concept is intrinsically broadband, and is straightforward to extend into the visible range to improve molecular sensing, single-photon sources, and non-linear light emitting devices.

**Results**

***Optical performance and fluorescence enhancement assessment.***

Our optical horn antenna platform addresses specifically the challenges of label-free single protein UV detection. It combines a reflective unit with a nanoaperture (Fig. 1a-c) and is the UV analogue of a microwave horn antenna (Supplementary Fig. S1). The central nanoaperture of 65 nm diameter concentrates the light in an attoliter detection volume to isolate a single protein and enhance its autofluorescence,[32,33] while the reflective conical unit covered with a 100 nm thick aluminum layer steers the autofluorescence light toward the microscope objective. Contrarily to the Yagi Uda or Bull's eyes resonant designs,[28,29] the conical horn is intrinsically broadband, covering the full 300-400 nm bandwidth, independently of resonance or interference effects. The detection volume provided by the 65 nm central aperture is three orders of magnitude below that of a diffraction-limited confocal microscope,[24,33] enabling single molecule detection at micromolar physiological conditions and circumventing the need for sub-nanomolar dilutions in conventional confocal microscopy.[34]

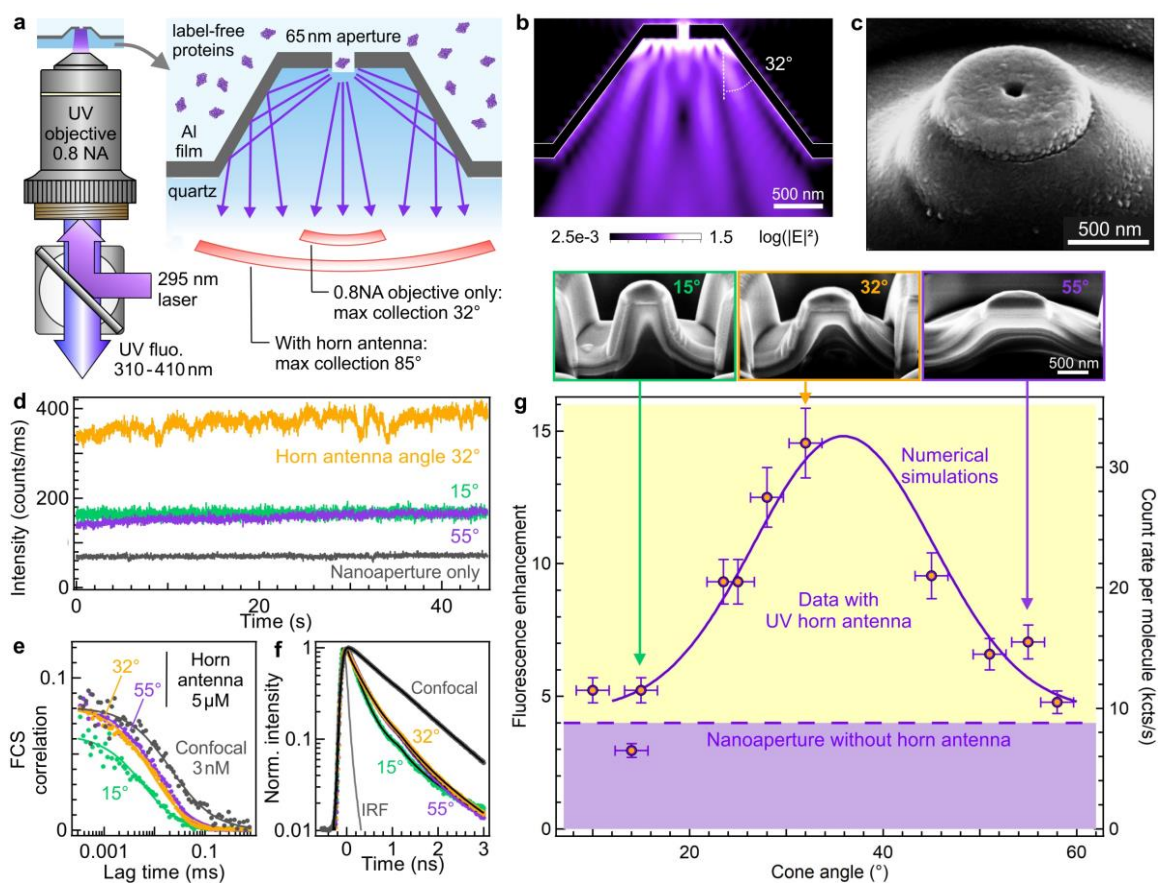

**Figure 1.** Ultraviolet horn antenna to enhance the autofluorescence detection of single label-free proteins. (a) Scheme of the experiment. (b) Numerical simulation of the emission pattern of a dipole located in the center of the nanoaperture, averaging the contributions from horizontal and vertical dipole orientations. (c) Scanning electron microscope image of a horn antenna. Similar images could be reproduced more than 10 times using the same milling parameters. (d) Fluorescence intensity time traces recorded on a 5 µM solution of p-terphenyl using horn antennas of different cone angles. (e) FCS correlation functions corresponding to the traces shown in (d), the case for the isolated nanoaperture is equivalent to the horn antennas with cone angles 32 and 55°. (f) Normalized fluorescence lifetime decay traces acquired simultaneously to the data in (d,e). IRF indicates the instrument response function. Black lines are numerical fits. (g) Fluorescence enhancement of the brightness per molecule as a function of the horn antenna cone angle. The right axis shows the corresponding count rate per p-terphenyl molecule at 80 µW of the 266 nm laser. The level achieved with a nanoaperture without any horn antenna is indicated by the dashed horizontal line. The solid line shows the numerical simulations results accounting for the collection efficiency gain into the 0.8 NA microscope objective. The SEM images in the inset show the antenna geometry after a cross-section has been cut by focused ion beam. Data are presented as mean values +/- one standard deviation determined from a pool of at least 3 different samples.

The horn antenna performance is assessed using p-terphenyl, a 93% quantum yield UV fluorescent dye (Fig. 1d-g). The conical reflector angle mainly determines the antenna collection efficiency into the 0.8 NA objective (Supplementary Fig. S2,S3). We have fabricated horn antennas with various cone angles from 10 to 55° (Supplementary Fig. S4-S6). The raw fluorescence time trace already shows a 4× larger signal with 32° horn as compared to the bare nanoaperture, indicating an improved collection efficiency by the same ratio (Fig. 1d). Fluorescence correlation spectroscopy (FCS) analysis measures the fluorescence brightness per molecule (Fig. 1e and Tab. S1) from which we compute the fluorescence enhancement as compared to the confocal reference.[24,29] The enhancement factor clearly depends on the cone angle with an optimum around 35° (Fig. 1g). While a single nanoaperture improves the p-terphenyl brightness by 4×,[33] the horn reflector brings it to 15×. These values are lower than previous reports using gold antennas in the red part of the spectrum,[22–25] yet this is explained by the ultraviolet range and the simple non-resonant design of the horn antenna. The major goal here is not to compete with plasmonics in the visible, but rather to enable UV autofluorescence detection of single proteins above the background noise. The collection efficiency gain has been numerically simulated using finite-difference time-domain (FDTD, solid curve in Fig. 1g). The simulation results match well with the experimental data trend, confirming the dependence with the cone angle.

The fluorescence lifetime measurements on Fig. 1f and Tab. S2 indicate that the p-terphenyl fluorescence lifetime is reduced by 3× in the antennas as compared to the confocal reference. This lifetime reduction is independent of the cone angle and is similar to the lifetime reduction found for the single aperture without the conical reflector. This shows that the emitter's fluorescence lifetime (and thus the local density of optical states) is mainly set by the aperture diameter. Having a similar local density of optical states between the nanostructured samples, we can conclude that the supplementary gain brought by the optimized horn antenna is directly related to the increase in directivity as compared to the bare nanoaperture. This confirms the idea of the conical reflector as a collection unit to steer the emitted light towards the microscope objective. Based on the gain respective to the bare nanoaperture, the maximum collection angle is estimated to be around 85° for our best system with 32° cone angle. Our nanophotonic platform collects the fluorescence light emitted at high angles, even beyond the supercritical angle. For a description of the aluminum nanoaperture influence on the fluorescence process, our group has recently published a detailed characterization using label-free proteins in the UV.[35] As shown by the data in Fig. 1, the presence of the horn reflector improves the collection efficiency, but the fluorescence excitation and emission enhancements occurring in the nanoaperture are not affected. We independently confirm the fluorescence enhancement by quantifying the noise reduction in the correlation data (Supplementary Fig. S7). The high brightness observed with the optimized horn antenna directly improves the FCS signal to noise ratio allowing to reduce the experiment integration time as compared to the confocal reference while keeping the same accuracy.

***Single immobilized protein detection.***

We next focus on the label-free detection of single immobilized proteins using our optimized UV horn antennas with 32° cone angle (Fig. 2). The surface of the central nanoaperture is functionalized with silane-polyethylene glycol-biotin to graft individual β-galactosidase-streptavidin proteins. β-galactosidase from Escherichia coli (156 tryptophans) has been modified to bear a streptavidin anchor (24 tryptophans). We first assess the distribution of the number of proteins inside the central nanoaperture using control experiments where the proteins are labelled with an Atto647N-biotin red fluorescent dye (Fig. 2a,b). Specific care was taken to ensure a 1:1 labeling ratio before the surface immobilization so that on average every protein carries a single fluorescent label (see Methods for details). The experiments are repeated for two protein concentrations of 5 and 0.1 nM, allowing to report the evolution with the protein concentration. With the Atto647N label, the fluorescence intensity time traces show fast step decays typical of single molecule fluorescence photobleaching (Fig. 2a,b). These traces are analyzed using the automated software AutoStepfinder [36] to count the number of bleaching steps from which we deduce the number of proteins. The histogram of the number of molecules per horn antenna follow the expected Poisson distribution, with an average number of molecules of 2.3 for the 5 nM concentration and 0.7 for the 0.1 nM case. The evolution of the average number of molecules does not scale exactly with the concentration (although there is an obvious dependence) as the 30 minutes incubation time and possible steric hindrance between neighboring proteins may limit the number of proteins able to bind on the horn antenna surface. Importantly, for the 0.1 nM protein concentration, there is less than one protein per horn antenna on average, so this case is representative of single molecule experiments.

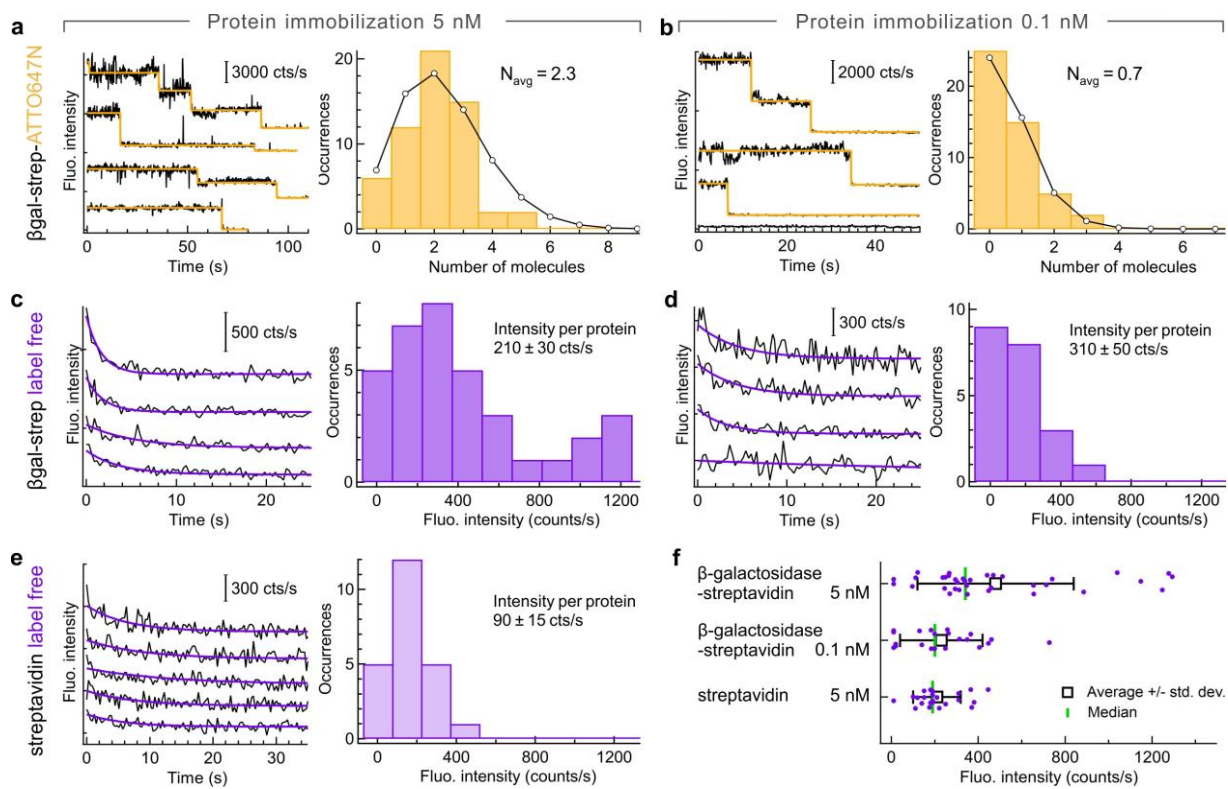

**Figure 2.** Horn antenna-enhanced label-free detection of immobilized single proteins. The surface of the horn antenna has been functionalized with biotin-PEG-silane to bind β-galactosidase-streptavidin and pure streptavidin proteins. In (a,c,e) the total protein concentration is 5 nM while it is reduced to 0.1 nM in (b,d). (a,b) Additional Atto647N-biotin red fluorescent label is added as a control to quantify the number of proteins inside individual horn antennas. Typical Atto647N fluorescence time traces are shown on the left panels in (a,b), together with their step-function fit (yellow line) allowing to count single molecules. The traces are vertically shifted for clarity. The right panels in (a,b) display the histogram of the number of detected molecules (bars) together with a Poisson distribution fit (line and dots). For (c-f) the experiments are performed with label-free proteins in the UV using 5 nM β-galactosidase-streptavidin (c), 0.1 nM β-galactosidase-streptavidin (d) and 5 nM streptavidin (e). In (c-e) the left panels show autofluorescence time traces and their exponential decay fit recorded on different individual horn antennas (the traces are vertically shifted for clarity). The right panels represent the histogram of the exponential fit amplitudes. For (c) 35 horn antennas were probed, for (d) 21 and (e) 23. (f) Scatter plot of the exponential fit amplitudes corresponding to the histograms in (c-e). The points are vertically shifted using a uniform statistical distribution to better view the results. The white square marker denotes the average value with the bars extending to one standard deviation. The green vertical line indicates the median. Respectively 35, 21 and 23 antennas were probed for the different cases from top to bottom. Throughout these experiments the cone angle is 32° and the aperture diameter is 200 nm (see Supplementary Fig. S13 for a discussion on the aperture diameter).

UV autofluorescence time traces are shown in Fig. 2c-e and S8-S10. These label-free experiments have been performed in exactly the same surface immobilization conditions as for Fig. 2a,b, so it is fair to consider that the distributions of the number of proteins are unchanged. Due to the high number of tryptophan residues in each protein, the autofluorescence time traces no longer bear step-like decays but follow an exponential decay due to photobleaching. Control experiments performed in the absence of the protein using only the photostabilizing buffer show that the background noise does not show the exponential signal decay upon UV illumination (Supplementary Fig. S11). In presence of the proteins, each trace stemming from a different horn antenna is fitted with an exponential function to extract the decay amplitude. The histograms of the autofluorescence signal amplitudes are displayed in Fig. 2c,d for 5 and 0.1 nM protein concentration, while the raw data, average, standard deviation and median values are summarized in Fig. 2f. Dividing the average fluorescence intensity (Fig. 2f) by the average number of proteins (determined from Fig. 2a,b), we estimate the average brightness per β-galactosidase-streptavidin protein in the horn antenna. We find 210 ± 30 counts/s for the 5 nM concentration and 310 ± 50 counts/s for the 0.1 nM concentration, which yield statistically quite comparable values (the difference is twice the standard deviation).

The experiments are reproduced with pure streptavidin, which has 24 tryptophan residues as compared to the 156 residues of β-galactosidase. A separate calibration using diffusing molecules shows that the average brightness per protein is 3× lower for streptavidin as compared to β-galactosidase.[35] Although streptavidin has a 6.5× lower absolute number of tryptophan residues, their average quantum yield is estimated to be around 3.5% while it is only of 1.6% in β-galactosidase due to a higher quenching by nearby aminoacids.[18,35] The experiments on immobilized streptavidin proteins are performed using a 5 nM concentration to work in the same conditions as for Fig. 2a,c. The average amplitude found for streptavidin is 2.5× lower than for β-galactosidase (Fig. 2e,f), confirming the expected evolution of the signal with the number of tryptophan residues and their average quantum yield. Finally, the experiments with immobilized proteins are reproduced for single nanoapertures without the horn microreflector (Supplementary Fig. S12). The signal is about 3× brighter with the horn antenna as compared to the nanoaperture on a flat substrate, which goes along with the results found for p-terphenyl in Fig. 1g and the improved collection performance of the horn antenna.

Altogether, the data presented in Fig. 2 and S8-12 demonstrate that the autofluorescence from a single protein can be recorded on a horn antenna. The signal scales with the protein concentration, the number of tryptophan residues and the collection efficiency. Experiments using a fluorescent marker allow an independent measurement of the number of proteins. This realizes label-free single protein detection in the UV. Additionally, fluorescence lifetime histograms can be extracted for traces corresponding to a single protein (Supplementary Fig. S14), a highly challenging task owing to the limited total photon budget that has to be distributed among the histogram time bins.

*Detecting single diffusing proteins.*

To confirm the single-molecule sensitivity claim, we perform experiments with diffusing β-galactosidase-streptavidin proteins at very low concentrations so that the average number of proteins present in the detection volume is significantly below 1 (Figure 3). For these experiments, we use a 200 nm diameter nanoaperture to increase the protein residency time inside the horn antenna. For the highest concentration of 20 nM, the calculated average number of proteins inside the 200 nm nanoaperture is 0.12,[33] which corresponds well to the regime required to observe fluorescence bursts from single molecules.[23,37] Moreover, we add 55% sucrose to the buffer solution to increase the viscosity and ensure the proteins stay a sufficiently long time of several milliseconds inside the nanoaperture volume. Without sucrose or any other viscous medium, the diffusion time of proteins across the nanoaperture would be below 1 ms which is not sufficient to record enough photons and clearly resolve the UV autofluorescence bursts stemming from a single protein. The presence of impurities in sucrose leads to a higher background noise level and limits the maximum amount of sucrose that we can use (glycerol mixtures lead to similar observations). We find that a 55% w/w sucrose mixture is a good compromise between increased viscosity and tolerable noise level.

Figure 3a shows typical autofluorescence time traces recorded with increasing protein concentrations from 0 (background only) to 20 nM. Without the β-galactosidase-streptavidin proteins, we detect 14 events per minute above the threshold of 310 counts per 30 ms bin time (corresponding to 2.5× the standard deviation of the background noise, dashed horizontal line in Fig. 3a). In the presence of the protein sample, the number of events exceeding this threshold increases with the protein concentration (Fig. 3a and 3b inset). Zooming in on some selected bursts (star markers in Fig. 3a), the signal is above the average for several binning times, which indicates that these bursts are not spurious single-time bin events. The photon count histograms confirm an increasing difference with the background level as the protein concentration is increased. Comparing the respective maxima, we estimate that the brightness for diffusing molecules is about 600 counts per second, which stands in agreement with an independent FCS calibration at micromolar concentration (Supplementary Fig. S15). To confirm that the bursts seen on the autofluorescence traces stem from the proteins and are not just random noise, we compute the temporal correlations of the traces in Fig. 3a and obtain the FCS correlograms on Fig. 3c. In the presence of the proteins, the FCS correlation is significantly higher than the residual background noise correlation. The FCS amplitude increases with the concentration because of the dominating presence of the background noise at such low concentrations.[38] The positive correlation amplitude and the 5 ms diffusion time (due to the presence of sucrose) indicate that the bursts seen on Fig. 3a stem from single β-galactosidase-streptavidin proteins. Altogether, the data in Fig. 3 demonstrate the ability of the UV horn antenna to resolve the autofluorescence bursts from diffusing single label-free proteins.

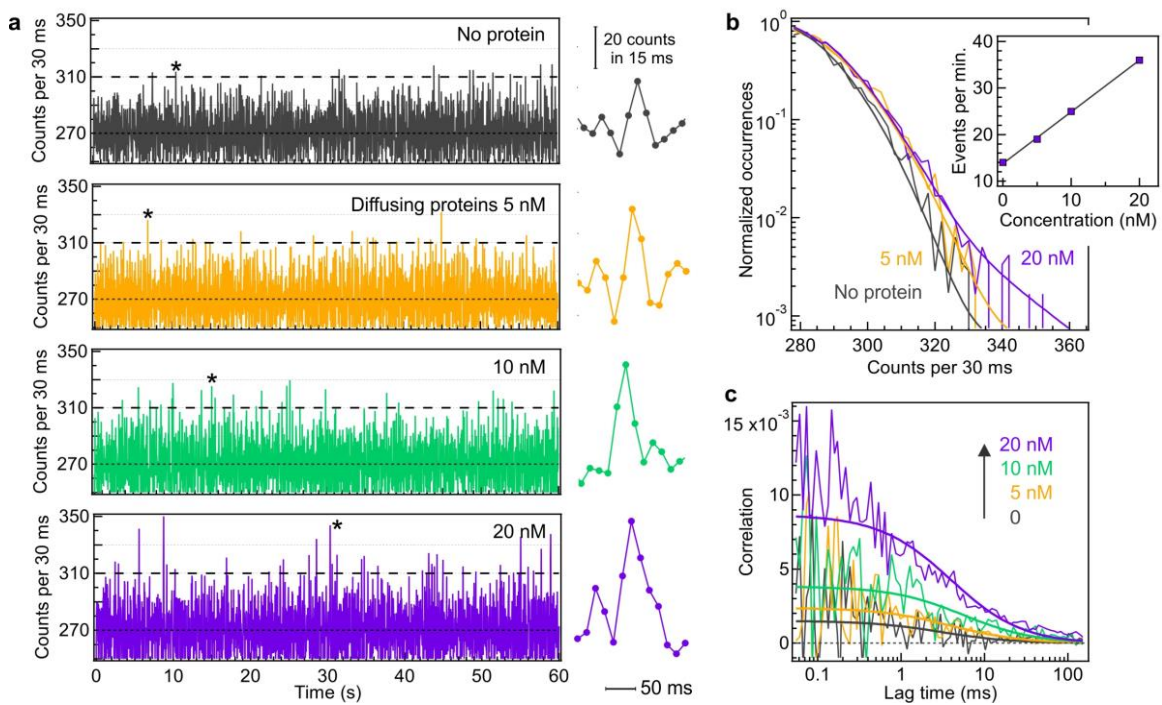

**Figure 3.** Label-free detection of single diffusing proteins across the UV horn antenna. 55% sucrose was added to the buffer to slow down the protein diffusion by 30× and ease observing the autofluorescence bursts. (a) Autofluorescence time traces with increasing β-galactosidase-streptavidin concentrations. The binning time is 30 ms. The stars indicate selected autofluorescence bursts which are displayed on the right panel with 15 ms bin time to better view individual bursts. (b) Normalized photon counts histograms computed over the full trace duration of 200 s. The thick lines are numerical fits using the sum of a Gaussian and an exponential distribution to account for the noise and the autofluorescence bursts respectively. The inset shows the number of detected events per minute above the threshold corresponding to 2.5× the standard deviation of the noise when no protein is present (dashed lines at 310 counts in (a)). (c) FCS correlation functions of the traces displayed in (a). The thick lines are numerical fits.

*Label-free protein denaturation and unfolding.*

β-galactosidase has gained importance as a model system to study protein folding.[39] Ensemble-level spectroscopy measurements such as the data in Fig. 4a,b and Supplementary Fig. S16 are commonly used to follow β-galactosidase denaturation in presence of urea.[40] It is believed that the tetrameric β-galactosidase first unfolds into a globular structure which then dissociates into unfolded monomers as the urea concentration is increased.[39,40] However, this pathway has never been studied at the single molecule level and only indirect ensemble-averaged measurements are available. Single molecule resolution is important here as it allows to clearly distinguish the dissociation into monomers by counting the number of molecules. An approach using fluorescence labelling would be very complicated as it would require all the β-galactosidase monomers to be labeled with a fluorescent dye, which is challenging to achieve and would lead to inter-chromophoric quenching due the close proximity of dyes.[41] Another motivation for single molecule resolution is to be able to simultaneously measure the protein hydrodynamic radius to clearly evidence unfolding.

Here we use label-free FCS enhanced by the optical horn antenna to study β-galactosidase denaturation (Fig. 4c,d). Without the horn antenna, the FCS data is too noisy to reliably measure unfolding and monomer dissociation (Supplementary Fig. S15). Thanks to the horn antenna, the UV autofluorescence brightness of β-galactosidase tetramers is increased by 10× which directly reduces the FCS noise by the same factor (Supplementary Fig. S15). In these conditions, the number of detected proteins and their hydrodynamic radius can be assessed as the urea concentration is increased (Fig. 4c,d and Supplementary Fig. S17). The number of molecules informs about the dissociation of the β-galactosidase tetramers into monomers while the hydrodynamic radius indicates expansion or compaction of the protein structure. As additional advantage, the attoliter detection volume of the horn antenna seems beneficial to avoid observing large aggregates which are a perturbation source for ensemble measurements.[40] While the autofluorescence spectrum of β-galactosidase is red-shifted up to 10 nm by the presence of urea (Fig. 4a and S16a,b), the total integrated intensity in the 310-410 nm detection range remains almost unaffected (Fig. S16c) as well as the autofluorescence lifetime (Fig. S16d).

Our experimental data show that below 3.5 M urea, β-galactosidase remains as a tetramer (the number of molecules is constant) while its hydrodynamic radius increases from 6.5 to 9.5 nm indicating protein unfolding (Fig. 4d). Our analysis takes into account the viscosity change as the urea concentration is increased (Supplementary Fig. S17) as well as the influence of the nanoaperture calibrated in [33]. In the absence of urea, our 6.5 ± 0.6 nm value for the hydrodynamic radius corresponds well to the 6.85 nm (experimental) and 6.7 nm (calculated) values determined previously in Ref.[42] Above 3.5 M urea, the hydrodynamic radius decreases down to 4 nm and the number of molecules increases by more than 2×. These are clear evidences of the tetramer partial dissociation into unfolded monomers, providing a confirmation of the β-galactosidase denaturation pathway established using ensemble methods.[39,40]

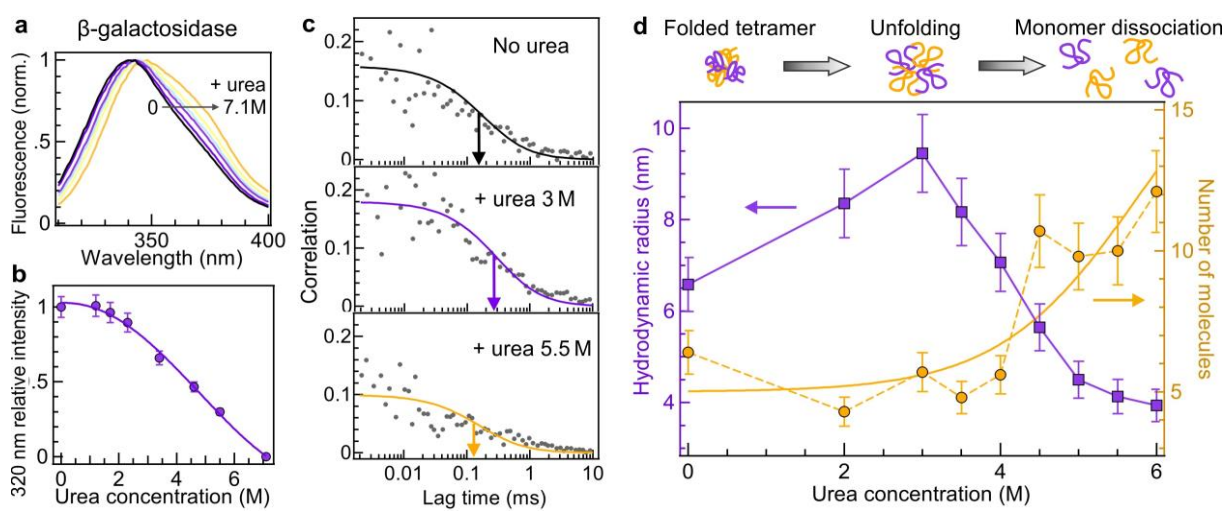

**Figure 4.** Application of UV horn antenna to study denaturation of label-free proteins. (a) Autofluorescence spectra of β-galactosidase in presence of increasing concentrations of urea. (b) Following the approach in [40], the autofluorescence intensity at 320 nm is used to monitor the denaturation of β-galactosidase as a function of the urea concentration. Data are presented as mean values, with the error bars corresponding to a 7% deviation accounting for the measurement uncertainty. The measurements have been repeated twice. (c) FCS correlation functions of β-galactosidase with increasing urea concentrations. The color lines are numerical fits. The arrows indicate the half-width of the correlation. (d) Evolution of the protein average hydrodynamic radius (left axis) and the mean number of detected molecules (right axis) as a function of the urea concentration. Error bars are standard deviations of the FCS measurements. Each measurement has been reproduced independently at least two times. The yellow line is a guide to the eyes based on a sigmoid function fit $5 + 15/(1 + e^{6-c})$ where $c$ is the urea concentration in molar.

**Discussion**

While label-free alternatives to fluorescence labelling are actively searched,[9–11] fluorescence spectroscopy remains by far the most widely used approach for single molecule detection. The optical horn antennas developed here make a significant step forward by enabling the direct detection of single label-free proteins via their natural ultraviolet fluorescence. This dedicated design combines simultaneously plasmonic fluorescence enhancement, efficient fluorescence collection, attoliter detection volume and strong background rejection, allowing to achieve unprecedented protein autofluorescence brightness. Improving the net detected UV photon count rate is key to enable the biophysical applications investigating single proteins in their native state in real time. Photodamage of aminoacids due to UV illumination is a potential issue that may affect the protein structure. For diffusing proteins, the short illumination time balances the negative impact of the UV photodegradation. However, for immobilized proteins, the risk of photodamage limits the maximum UV power exciting the protein. We have used so far the lowest possible power of 0.3 µW, yet this phenomenon is currently setting the limit for the achievable signal to noise ratio. UV detection of single molecules is still at its infancy, and we hope that this work will stimulate more studies in this direction. Moreover, optical horn antennas are also beneficial to improve the collection efficiency in the visible regime,[28,29] and analyze single fluorescent molecules in a crowded and confined environment reproducing the physiological conditions.[34]

**Methods**

*Optical horn antenna fabrication*

The fabrication process involves several steps (Supplementary Fig. S4). Briefly, we first mill the horn antenna by focused ion beam (FIB). Then a 100 nm aluminum layer is deposited to ensure a good UV reflectivity of the horn antenna walls. Lastly we carve a 65 nm or a 200 nm diameter nanoaperture by FIB in the center of the top plateau of the horn antenna. The substrates are cleaned NEGS1 quartz microscope coverslips of 150 µm thickness (Neyco). Aluminum layers are deposited by electron-beam evaporation (Bühler Syrus Pro 710) with 10 nm/s rate at a chamber pressure of $10^{-6}$ mbar. FIB milling is performed using a gallium-based system (FEI dual beam DB235 Strata) with 30 kV acceleration voltage and 300 pA current for milling the horn antenna and 10 pA current for milling the central nanoaperture. All nanoapertures have a 50 nm deep undercut into the quartz substrate to maximize the signal enhancement. Lastly, a 12 nm-thick $SiO_2$ layer is deposited by plasma-enhanced chemical vapor protection (PECVD, PlasmaPro NGP80 from Oxford Instruments) to protect the aluminum surface against corrosion.[43,44]

*Protein samples and photostabilizing buffer*

β-galactosidase from Escherichia coli (156 tryptophan residues, PDB 1DP0), β-galactosidase-streptavidin conjugate (180 tryptophan residues) and streptavidin from Streptomyces avidinii (24 tryptophan residues) are purchased from Sigma-Aldrich. The proteins are dissolved in a Hepes buffer (25 mM Hepes, 300 mM NaCl, 0.1 v/v% Tween20, 1 mM DTT and 1 mM EDTA 1mM at pH 6.1) which was reported to stabilize β-galactosidase conformation and avoid aggregate formation.[40] All the protein stock solutions have been centrifuged for 12 min at 142,000 g (Airfuge). Just before the optical measurements, GODCAT oxygen scavenger (100 nM glucose oxidase, 830 nM catalase, 10 w/v% D-glucose) with 10 mM DABCO (1,4-Diazabicyclo[2.2.2]octane) is added to the solution to improve the UV photostability.[18] p-terphenyl is also used as received from Sigma-Aldrich and diluted in HPLC-grade cyclohexane.

*Surface immobilization of proteins*

The horn antennas are thoroughly cleaned with UV-ozone (Novascan PSD-UV 100 W), ethanol rinsing and oxygen-plasma (Diener Zepto 50W, 0.6 mbar, 10 min). The silica surface is then coated with silane-modified polyethylene glycol (PEG-silane) by immersion into an ethanol solution containing 0.75 mg PEG-silane 1000 Da and 1.2 mg biotin-PEG-silane 2000 Da. PEG-silane and biotin-PEG-silane are purchased from Nanocs. After overnight incubation, the sample is rinsed with ethanol and dried. The protein solution with either 5 or 0.1 nM concentration is placed on the surface for 30 min allowing the streptavidin anchor to bind the surface-grafted biotin. Before the optical measurements, the surface is rinsed three times with water and covered with the GODCAT photostabilizing buffer.

*Control experiments with fluorescent labels*

A solution of β-galactosidase-streptavidin is mixed with a solution of Atto647N-biotin in a 1:1 ratio and kept at 4°C overnight. The concentrations were carefully checked using a spectrofluorometer (Tecan Spark 10M) to ensure proper labelling of the proteins with a single Atto647N dye. In this configuration, on average each protein is expected to bind only one fluorescent tag via the streptavidin-biotin bridge. Assuming that each β-galactosidase-streptavidin has 3 binding sites for the biotinylated fluorescent tag (one binding site is taken by the anchor between β-galactosidase and streptavidin), the probability that a protein bears more than 1 fluorescent label is estimated to be below 4%. The surface grafting of the labeled β-galactosidase-streptavidin is performed using exactly the same protocol used for the label-free counterpart. The fluorescence readout is performed on a confocal microscope described in ref.[26] using 1 µW laser power at 635 nm. To promote Atto647N photostability and minimize blinking, we use the GODCAT oxygen scavenger system together with 1 mM Trolox ((±)-6-Hydroxy-2,5,7,8-tetramethylchromane-2-carboxylic acid).[45] Prior to the experiments, the Trolox stock solution in DMSO is illuminated with a UV lamp for 25 min to ensure a proper ratio of Trolox and its Trolox-quinone derivative.[45]

*Urea denaturation*

A 1.7 µM β-galactosidase protein solution is incubated with urea at various concentrations from 0 to 6 M (pH 7) for 90 minutes at room temperature. 0.4 v/v% Tween 20 is added to the buffer to minimize aggregation of denaturated proteins. To avoid non-specific adsorption of proteins, the horn antenna surface is passivated with PEG-silane 1000 Da (Nanocs) by immersion into a 1 mg/ml PEG-silane solution ethanol with 1% acetic acid for 3-4 hours followed by ethanol rinsing. GODCAT photostabilizing system is added to the protein buffer just before the optical measurements and the urea concentration is adjusted to keep a constant value to avoid protein refolding.

*UV microscope*

We operate a custom-built confocal microscope with a LOMO 58x, 0.8 NA, water immersion objective. Experiments on p-terphenyl use a 266 nm picosecond laser (Picoquant LDH-P-FA-266, 70 ps pulse duration, 80 MHz repetition rate) with 80 µW average power, while experiments on proteins use a 295 nm picosecond laser (Picoquant VisUV-295-590, 70 ps pulse duration, 80 MHz repetition rate). The 295 nm wavelength selectively excites tryptophan residues, as tyrosine and phenylalanine have negligible absorption above 290 nm. The laser power for immobilized protein detection is 0.3 µW, while for diffusing molecules we use 8 µW. The urea denaturation experiments on diffusing proteins are performed at 10 µW. Both laser beams are spatially filtered to ensure a Gaussian profile filling the objective back aperture, they pass through a short pass filter (Semrock FF01-311/SP-25) and are reflected by a dichroic mirror (Semrock FF310-Di01-25-D). The optical horn antenna is positioned at the laser focus with a 3-axis piezoelectric stage (Physik Instrumente P-517.3CD). For the immobilized protein experiment, we use the microscope LED illumination to localize the horn antenna. The laser illumination is turned on immediately at the start of the acquisition to avoid bleaching the proteins while scanning the sample.

The fluorescence light is collected back by the microscope objective and separated from the laser light by the dichroic mirror and two emission filters (Semrock FF01-300/LP-25 and Semrock FF01-375/110-25). The spectral range for fluorescence detection goes from 310 to 410 nm. Confocal detection is performed using a 200 mm focal length doublet lens (Thorlabs ACA254-200-UV) and a 80 µm pinhole. Single photon counting uses a photomultiplier tube (Picoquant PMA 175) connected to a photon counting module (Picoquant Picoharp 300 with time tagged time resolved mode). The integration time is 2 to 3 minutes.

*Fluorescence time trace analysis*

The fluorescence data is analyzed with Symphotime 64 (Picoquant) and Igor Pro 7 (Wavemetrics). For the fluorescence time traces from immobilized label-free single proteins, the temporal bin width is set to 300 ms to optimize the signal to noise ratio while still providing enough time resolution. Each trace stemming from a different horn antenna is fitted using and exponential decay model $A\, e^{-t/\tau_B} + y_0$ where $A$ is the decay amplitude (used for the histrograms), $\tau_B$ is the bleaching time and $y_0$ is the background level set by the dark counts of the photodetector and the residual photoluminescence from the nanostructure. For the control experiments using Atto647N labels, the traces are analysed using AutoStepfinder, a recently developed software for automatic step detection.[36]

For the experiments on diffusing proteins at very low concentrations (average numbers of proteins in the nanoaperture < 0.12, Fig. 3), we apply a 1 Hz high-pass filter to remove all long-term drifts and fluctuations. These drifts have a small amplitude (typically 5 to 10 counts per 30 ms) and long periods (several seconds), yet due to the low signal to noise ratio at low protein concentrations, the long-term fluctuations have to be removed before computing the photon count histogram. The average value of the initial trace is then added to retrieve the filtered count information.

FCS correlations are fitted using a three dimensional Brownian diffusion model with a blinking term:[18,33]

$$G(\tau) = \frac{1}{N_{\text{mol}}} \left[1 - \frac{B}{F}\right]^2 \left(1 + n_T \exp\left(-\frac{\tau}{\tau_T}\right)\right) \left(1 + \frac{\tau}{\tau_d}\right)^{-1} \left(1 + \frac{1}{\kappa^2}\frac{\tau}{\tau_d}\right)^{-0.5} \quad (1)$$

where $N_{\text{mol}}$ is the total number of detected molecules, $B$ the background noise intensity, $F$ the total fluorescence intensity, $n_T$ and $\tau_T$ are the blinking amplitude and characteristic time, $\tau_d$ is the mean diffusion time and κ the aspect ratio of the axial to transversal dimensions of the detection volume (κ

= 8 for the confocal illumination and κ = 1 for the horn antenna). Note that the fast kinetics components determined by $n_T$ and $\tau_T$ may not be only related to triplet blinking, they could also account for residual afterpulsing from the photon counting detector and/or metal quenching when the protein diffuses in nanometer proximity to the aluminum surface. The FCS fit results are summarized in the supplementary information.

The fluorescence lifetime decays are fitted by an iterative reconvolution taking into account the measured instrument response function (IRF). As noted in our previous work on single apertures,[33] a three exponential model is needed to correctly interpolate the experimental data. A fixed 10 ps component accounts for the metal photoluminescence and Raman scattering background, while a long component with a lifetime similar to the confocal reference corresponds to a residual fluorescence stemming from molecules away from the optical antenna. The intermediate lifetime component (which has a dominating intensity) corresponds to the lifetime of molecules inside the horn antenna. All the fit results are detailed in the supplementary information.

*Numerical simulations*

The electric field intensity radiated by a point dipole is calculated with finite-difference time-domain (FDTD) method using RSoft Fullwave software. We set a fixed emission wavelength at 350 nm. Horizontal and vertical orientations of the source dipole are computed separately and averaged for the final output to represent the emission of a molecule with nanosecond rotational time. Each simulation is run with 2 nm mesh size and is checked for convergence after several optical periods. To compute the enhancement displayed on Fig. 1g, 10 horn antennas with different cone angle are simulated. For each of them, we compute the gain into the 0.8 collection NA of the microscope objective as compared to the emission from a single aperture without horn antenna (Fig. S2). The data points are then fitted with a Gaussian function.

**Supplementary Information**

Microwave horn antenna analogy, FDTD simulations of dipole emission, Horn antenna fabrication, FCS and lifetime data tables for p-terphenyl, FCS noise analysis, Additional autofluorescence time traces, Background when no protein is present, Comparison of single protein autofluorescence time traces with and without horn antenna, Dependence with the aperture diameter, Fluorescence lifetime measurements of single label-free proteins, Fluorescence enhancement of β-galactosidase with horn antennas, β-galactosidase autofluorescence spectra in presence of urea, Diffusion time of β-galactosidase and viscosity calibration in presence of urea

**Data availability**

All relevant data are available from the corresponding author upon request.

**References**

1.  Joo, C., Balci, H., Ishitsuka, Y., Buranachai, C. & Ha, T. Advances in Single-Molecule Fluorescence Methods for Molecular Biology. *Annu. Rev. Biochem.* **77**, 51–76 (2008).

2.  Lerner, E. *et al.* Toward dynamic structural biology: Two decades of single-molecule Förster resonance energy transfer. *Science* **359**, eaan1133 (2018).

3.  Cabantous, S., Terwilliger, T. C. & Waldo, G. S. Protein tagging and detection with engineered self-assembling fragments of green fluorescent protein. *Nat. Biotechnol.* **23**, 102-107 (2005).

4.  Wigley, W. C., Stidham, R. D., Smith, N. M., Hunt, J. F. & Thomas, P. J. Protein solubility and folding monitored in vivo by structural complementation of a genetic marker protein. *Nat. Biotechnol.* **19**, 131–136 (2001).

5.  Dietz, M. S., Wehrheim, S. S., Harwardt, M.-L. I. E., Niemann, H. H. & Heilemann, M. Competitive Binding Study Revealing the Influence of Fluorophore Labels on Biomolecular Interactions. *Nano Lett.* **19**, 8245–8249 (2019).

6.  Riback, J. A. *et al.* Commonly used FRET fluorophores promote collapse of an otherwise disordered protein. *Proc. Natl. Acad. Sci.* **116**, 8889–8894 (2019).

7.  Zanetti-Domingues, L. C., Tynan, C. J., Rolfe, D. J., Clarke, D. T. & Martin-Fernandez, M. Hydrophobic Fluorescent Probes Introduce Artifacts into Single Molecule Tracking Experiments Due to Non-Specific Binding. *PLOS ONE* **8**, e74200 (2013).

8.  Hughes, L. D., Rawle, R. J. & Boxer, S. G. Choose Your Label Wisely: Water-Soluble Fluorophores Often Interact with Lipid Bilayers. *PLOS ONE* **9**, e87649 (2014).

9. Arroyo, J. O. & Kukura, P. Non-fluorescent schemes for single-molecule detection, imaging and spectroscopy. *Nat. Photonics* **10**, 11–17 (2016).

10. Zijlstra, P., Paulo, P. M. R. & Orrit, M. Optical detection of single non-absorbing molecules using the surface plasmon resonance of a gold nanorod. *Nat. Nanotechnol.* **7**, 379–382 (2012).

11. Hanay, M. S. *et al.* Single-protein nanomechanical mass spectrometry in real time. *Nat. Nanotechnol.* **7**, 602–608 (2012).

12. Baaske, M. D., Neu, P. S. & Orrit, M. Label-Free Plasmonic Detection of Untethered Nanometer-Sized Brownian Particles. *ACS Nano* **14**, 14212–14218 (2020).

13. Garoli, D., Yamazaki, H., Maccaferri, N. & Wanunu, M. Plasmonic Nanopores for Single-Molecule Detection and Manipulation: Toward Sequencing Applications. *Nano Lett.* **19**, 7553–7562 (2019).

14. Kumamoto, Y., Taguchi, A. & Kawata, S. Deep-Ultraviolet Biomolecular Imaging and Analysis. *Adv. Opt. Mater.* **7**, 1801099 (2019).

15. Lakowicz, J. R. *Principles of Fluorescence Spectroscopy*. (Springer US, 2006).

16. Li, Q. & Seeger, S. Label-Free Detection of Single Protein Molecules Using Deep UV Fluorescence Lifetime Microscopy. *Anal. Chem.* **78**, 2732–2737 (2006).

17. Lippitz, M., Erker, W., Decker, H., Holde, K. E. van & Basché, T. Two-photon excitation microscopy of tryptophan-containing proteins. *Proc. Natl. Acad. Sci.* **99**, 2772–2777 (2002).

18. Barulin, A. & Wenger, J. Ultraviolet Photostability Improvement for Autofluorescence Correlation Spectroscopy on Label-Free Proteins. *J. Phys. Chem. Lett.* **11**, 2027–2035 (2020).

19. Novotny, L. & Hecht, B. *Principles of Nano-Optics*. (Cambridge University Press, 2012).

20. Ruckstuhl, T., Enderlein, J., Jung, S. & Seeger, S. Forbidden Light Detection from Single Molecules. *Anal. Chem.* **72**, 2117–2123 (2000).

21. Novotny, L. & Hulst, N. van. Antennas for light. *Nat. Photonics* **5**, 83–90 (2011).

22. Kinkhabwala, A. *et al.* Large single-molecule fluorescence enhancements produced by a bowtie nanoantenna. *Nat. Photonics* **3**, 654–657 (2009).

23. Acuna, G. P. *et al.* Fluorescence Enhancement at Docking Sites of DNA-Directed Self-Assembled Nanoantennas. *Science* **338**, 506–510 (2012).

24. Punj, D. *et al.* A plasmonic 'antenna-in-box' platform for enhanced single-molecule analysis at micromolar concentrations. *Nat. Nanotechnol.* **8**, 512–516 (2013).

25. Puchkova, A. *et al.* DNA Origami Nanoantennas with over 5000-fold Fluorescence Enhancement and Single-Molecule Detection at 25 µM. *Nano Lett.* **15**, 8354–8359 (2015).

26. Flauraud, V. *et al.* In-Plane Plasmonic Antenna Arrays with Surface Nanogaps for Giant Fluorescence Enhancement. *Nano Lett.* **17**, 1703–1710 (2017).

27. Khatua, S. *et al.* Resonant Plasmonic Enhancement of Single-Molecule Fluorescence by Individual Gold Nanorods. *ACS Nano* **8**, 4440–4449 (2014).

28. Curto, A. G. *et al.* Unidirectional Emission of a Quantum Dot Coupled to a Nanoantenna. *Science* **329**, 930–933 (2010).

29. Aouani, H. *et al.* Bright Unidirectional Fluorescence Emission of Molecules in a Nanoaperture with Plasmonic Corrugations. *Nano Lett.* **11**, 637–644 (2011).

30. Lee, K. G. *et al.* A planar dielectric antenna for directional single-photon emission and near-unity collection efficiency. *Nat. Photonics* **5**, 166–169 (2011).

31. Morozov, S., Gaio, M., Maier, S. A. & Sapienza, R. Metal–Dielectric Parabolic Antenna for Directing Single Photons. *Nano Lett.* **18**, 3060–3065 (2018).

32. Levene, M. J. *et al.* Zero-Mode Waveguides for Single-Molecule Analysis at High Concentrations. *Science* **299**, 682–686 (2003).

33. Barulin, A., Claude, J.-B., Patra, S., Bonod, N. & Wenger, J. Deep Ultraviolet Plasmonic Enhancement of Single Protein Autofluorescence in Zero-Mode Waveguides. *Nano Lett.* **19**, 7434–7442 (2019).

34. Grommet, A. B., Feller, M. & Klajn, R. Chemical reactivity under nanoconfinement. *Nat. Nanotechnol.* **15**, 256–271 (2020).

35. Barulin, A., Roy, P., Claude, J.-B. & Wenger, J. Purcell radiative rate enhancement of label-free proteins with ultraviolet aluminum plasmonics. *J. Phys. Appl. Phys.* **54**, 425101 (2021).

36. Loeff, L., Kerssemakers, J. W. J., Joo, C. & Dekker, C. AutoStepfinder: A fast and automated step detection method for single-molecule analysis. *Patterns* **2**, 100256 (2021).

37. Baibakov, M. *et al.* Extending Single-Molecule Förster Resonance Energy Transfer (FRET) Range beyond 10 Nanometers in Zero-Mode Waveguides. *ACS Nano* **13**, 8469–8480 (2019).

38. Rüttinger, S. *Confocal Microscopy and Quantitative Single Molecule Techniques for Metrology in Molecular Medicine*. (Technischen Universität Berlin, 2006).

39. Juers, D. H., Matthews, B. W. & Huber, R. E. LacZ β-galactosidase: Structure and function of an enzyme of historical and molecular biological importance. *Protein Sci.* **21**, 1792–1807 (2012).

40. Nichtl, A., Buchner, J., Jaenicke, R., Rudolph, R. & Scheibel, T. Folding and association of β-galactosidase. *J. Mol. Biol.* **282**, 1083–1091 (1998).

41. Schröder, T., Scheible, M. B., Steiner, F., Vogelsang, J. & Tinnefeld, P. Interchromophoric Interactions Determine the Maximum Brightness Density in DNA Origami Structures. *Nano Lett.* **19**, 1275–1281 (2019).

42. Fleming, P. J. & Fleming, K. G. HullRad: Fast Calculations of Folded and Disordered Protein and Nucleic Acid Hydrodynamic Properties. *Biophys. J.* **114**, 856–869 (2018).

43. Barulin, A. *et al.* Preventing Aluminum Photocorrosion for Ultraviolet Plasmonics. *J. Phys. Chem. Lett.* **10**, 5700–5707 (2019).

44. Roy, P. *et al.* Preventing Corrosion of Aluminum Metal with Nanometer-Thick Films of Al2O3 Capped with TiO2 for Ultraviolet Plasmonics. *ACS Appl. Nano Mater.* **4**, 7199–7205 (2021).

45. Cordes, T., Vogelsang, J. & Tinnefeld, P. On the Mechanism of Trolox as Antiblinking and Antibleaching Reagent. *J. Am. Chem. Soc.* **131**, 5018–5019 (2009).

**Acknowledgments**

The authors thank Antonin Moreau, Julien Lumeau, Satyajit Patra, Manos Mavrakis, Redha Abdeddaim and Marco Abbarchi for stimulating discussions and technical help. This project has received funding from the European Research Council (ERC) under the European Union's Horizon 2020 research and innovation programme (grant agreement No 723241, A. B., P. R., J.-B; C., J. W.).

**Author contributions**

AB performed and analysed measurements, PR performed and analysed measurements and simulations, JBC performed nanofabrication, JW designed and supervised research and prepared manuscript with the help of all authors.

**Note**

The authors declare no competing interests.

**Supplementary Information for**

**Ultraviolet optical horn antennas for label-free detection of single proteins**

Aleksandr Barulin,[1] Prithu Roy,[1] Jean-Benoît Claude,[1] Jérôme Wenger[1,*]

[1] *Aix Marseille Univ, CNRS, Centrale Marseille, Institut Fresnel, 13013 Marseille, France*

* *Corresponding author:* jerome.wenger@fresnel.fr

**Contents:**

## S1. Microwave horn antenna analogy

**a** Microwave horn antenna

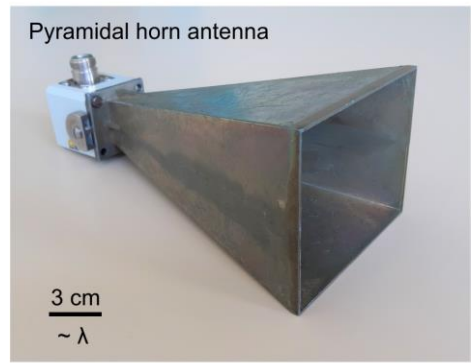
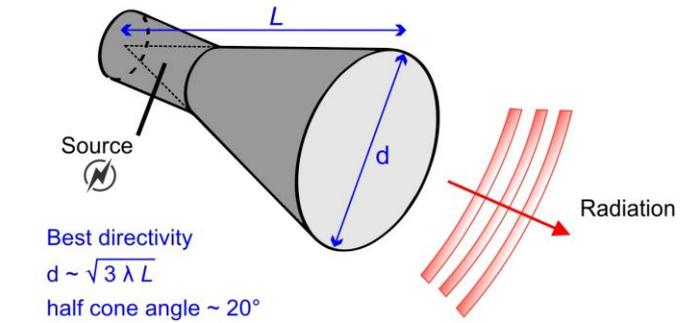

**b** Optical horn antenna

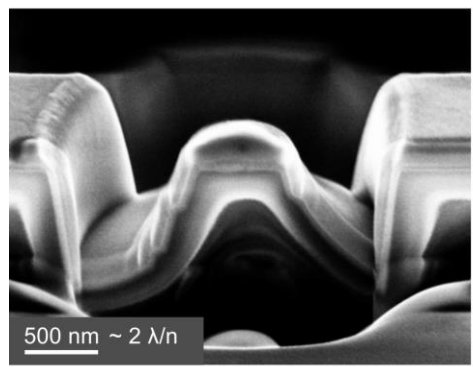
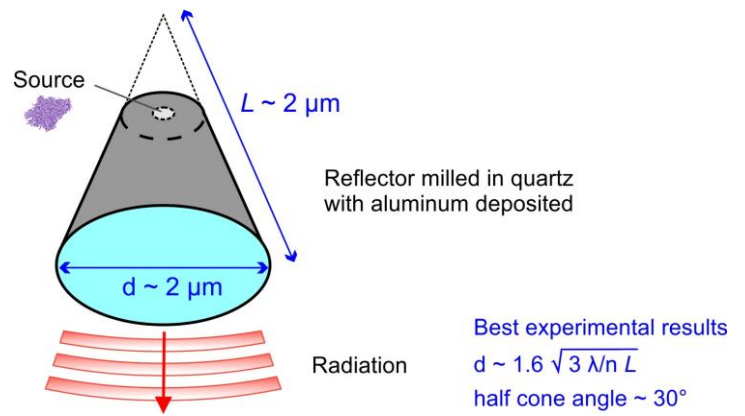

**Figure S1.** Analogy between microwave horn antennas (a) and optical horn antennas (b). Due to the circular symmetry of optical components (including the microscope objective), the conical horn design is most adapted. For microwave horn antennas, the directivity is optimized for a specific relationship between the antenna dimensions and the wavelength.[1] For our optical horn antennas, we retrieve a quite similar relationship, yet the large 310-410 nm spectral bandwidth in optics, the difficulty to control accurately the horn shape at the nanometer scale and the presence of resonances complicate the direct extrapolation of the microwave design formulas.[2]

## S2. FDTD simulations of dipole emission

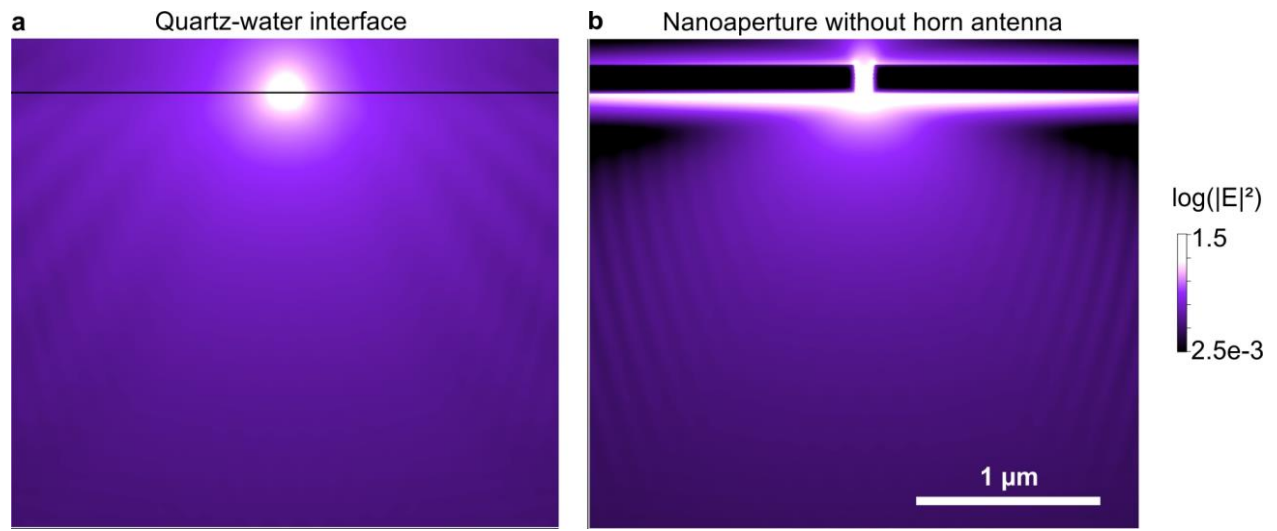

**Figure S2.** Finite difference time domain (FDTD) simulation of the emission pattern of a dipole located 10 nm above the quartz interface (a) and in presence of a 65 nm aluminum nanoaperture (b). The simulations are performed in the same conditions as in Fig. 1b, the colorscales are identical. In the absence of the horn antenna, the emission lies largely outside the 33° maximum collection angle of the 0.8 NA microscope objective.

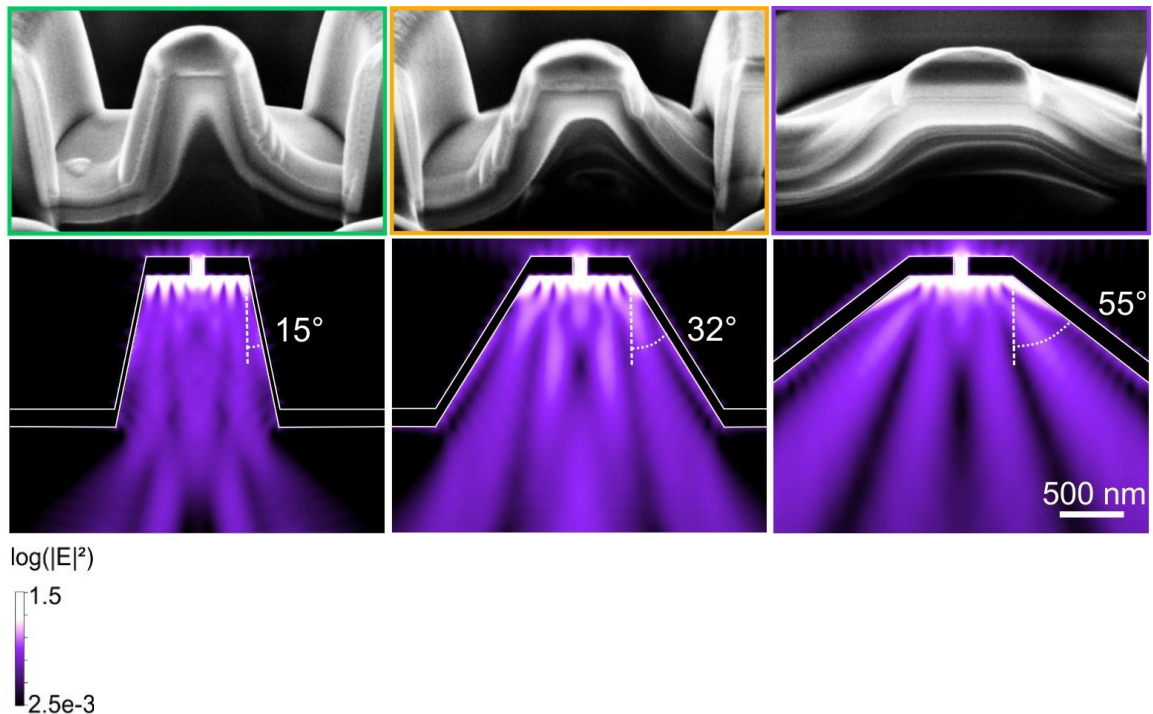

**Figure S3.** FDTD simulation of the emission pattern of a dipole located in the center of the 65 nm aperture, 10 nm above the quartz interface in presence of the conical horn antenna with different angles. The selected geometries reproduce the experimental horn antennas used in Fig. 1d,e. The colorscale is identical to Fig. 1b and Fig. S2. The contributions from horizontal and vertical dipole orientations are averaged.

## S3. Horn antenna fabrication

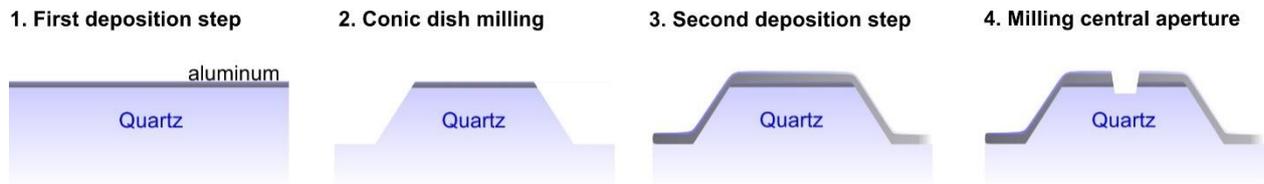

**Figure S4.** Fabrication protocol of the horn antenna platform: first the horn antenna is milled by focused ion beam (FIB) on a quartz coverslip coated with 50 nm aluminum layer to ensure a proper electrical conductivity (steps 1 and 2). Then a 100 nm aluminum layer is deposited on top of the horn antenna to make the horn antenna walls reflective in the UV (step 3). Finally a 65 nm diameter nanoaperture is milled in the center of the horn antenna top plateau (step 4).

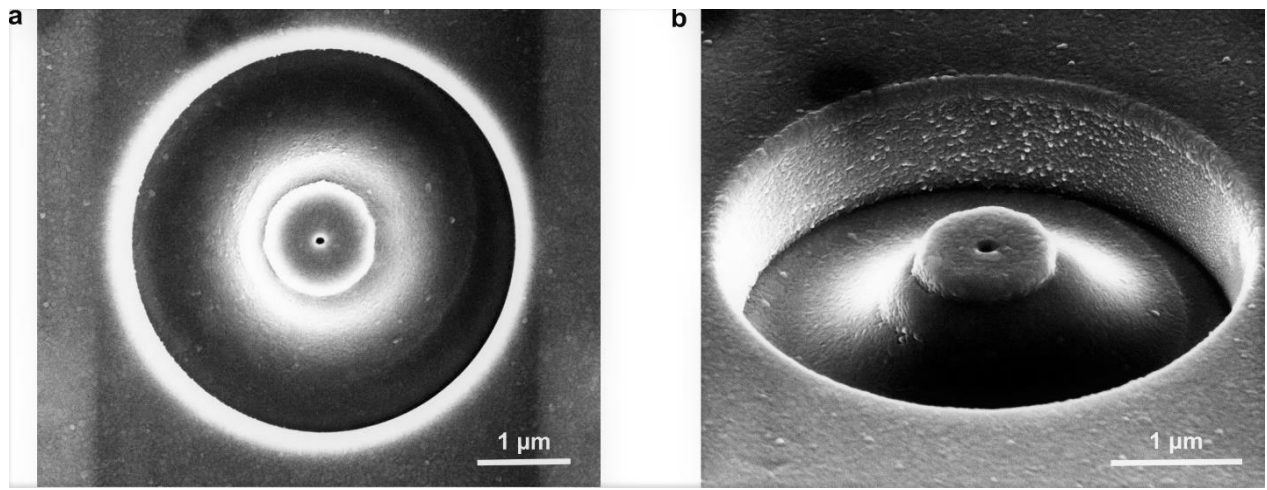

**Figure S5.** Scanning electron microscopy (SEM) images taken with normal incidence (a) and 52° tilt (b) of a complete horn antenna including the central 65 nm diameter nanoaperture. Similar images could be reproduced more than 10 times using the same milling parameters.

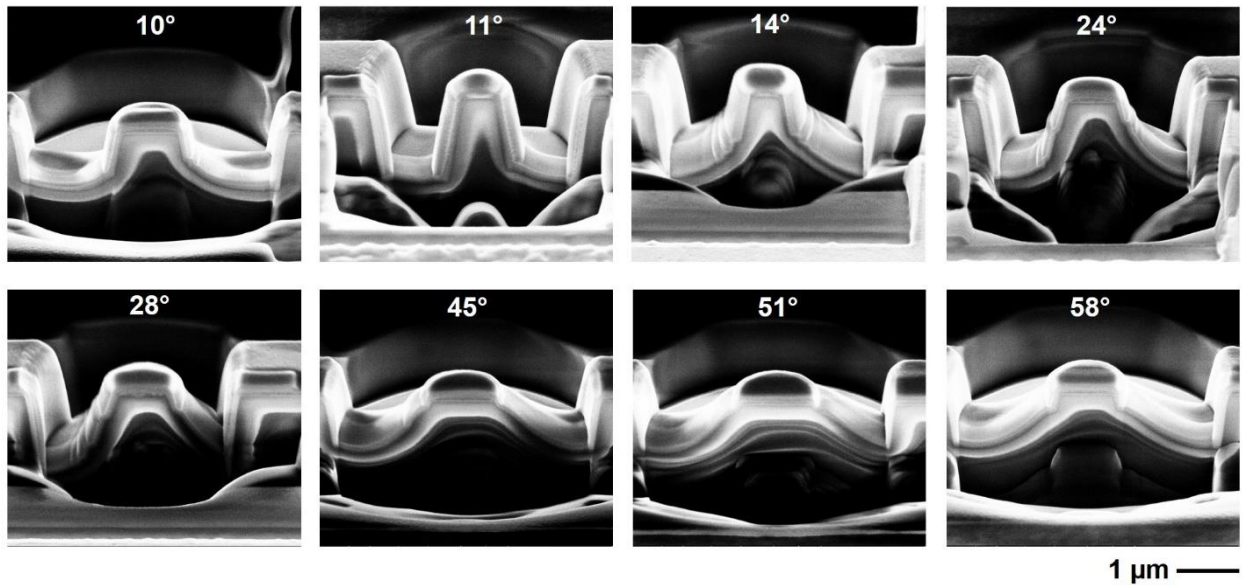

**Figure S6.** SEM images of different horn antennas with increasing cone angles as indicated on each picture. The sample is tilted by 52° and a FIB cross-section is performed to enable viewing the geometry of the device. The central nanoaperture has not been milled on these samples. We have checked 5 different independent samples leading to similar images.

## S4. FCS and lifetime data tables for p-terphenyl

**Table S1.** Fit parameters for the FCS data displayed on Fig. 1d,e. We do not observe any fast blinking contribution for p-terphenyl, so $n_T$ and $\tau_T$ are set to zero. The shape parameter κ is fixed at 8 for the confocal case based on the calibration of the microscope point spread function (PSF). For the aperture and horn antennas, we use κ = 1 based on our previous work using nanoapertures and fluorescent dyes in the visible.[3] *CRM = (F-B)/N$_{mol}$* stands for the fluorescence count rate per molecule (average brightness per emitter). A similar reduction in the translational diffusion time was observed for nanoapertures in the visible spectral range as compared to the diffraction-limited confocal reference.[3]

|  | *F* (kHz) | *B* (kHz) | *N*$_{mol}$ | τ$_d$ (μs) | *CRM* (kHz) | Fluo. enhancement |
|---|---|---|---|---|---|---|
| Confocal | 27.3 | 0 | 12.3 | 21.1 | 2.2 | - |
| Single aperture | 62.1 | 3.8 | 6.6 | 15.5 | 8.8 | 4 |
| Horn antenna 32° | 347.9 | 32 | 9.9 | 13.8 | 31.9 | 14.5 |
| Horn antenna 15° | 145 | 32 | 9.7 | 9.2 | 11.7 | 5.3 |
| Horn antenna 55° | 146.9 | 32 | 7.3 | 16.6 | 15.7 | 7.1 |

**Table S2.** Fit parameters for the fluorescence lifetime data displayed on Fig. 1f. The lifetimes are expressed in ns, the intensities are normalized so that their sum equals 1. All horn antennas and the single aperture share similar lifetime reductions. The horn antenna essentially affects the fluorescence collection, the local density of optical states is determined by the central 65 nm diameter aperture which remains constant among the different nanostructures.

|  | τ$_1$ | τ$_2$ | τ$_3$ | I$_1$ | I$_2$ | I$_3$ | Lifetime reduction (0.95 ns / τ$_2$) |
|---|---|---|---|---|---|---|---|
| Confocal | - | 0.95 | - | - | 1 | - | - |
| Single aperture | 0.01 | 0.32 | 0.95 | 0.23 | 0.47 | 0.3 | 3 |
| Horn antenna 32° | 0.01 | 0.34 | 0.95 | 0.24 | 0.35 | 0.41 | 2.8 |
| Horn antenna 15° | 0.01 | 0.25 | 0.95 | 0.31 | 0.4 | 0.29 | 3.8 |
| Horn antenna 55° | 0.01 | 0.35 | 0.95 | 0.19 | 0.49 | 0.32 | 2.7 |

## S5. The reduction of the FCS noise confirms the fluorescence enhancement

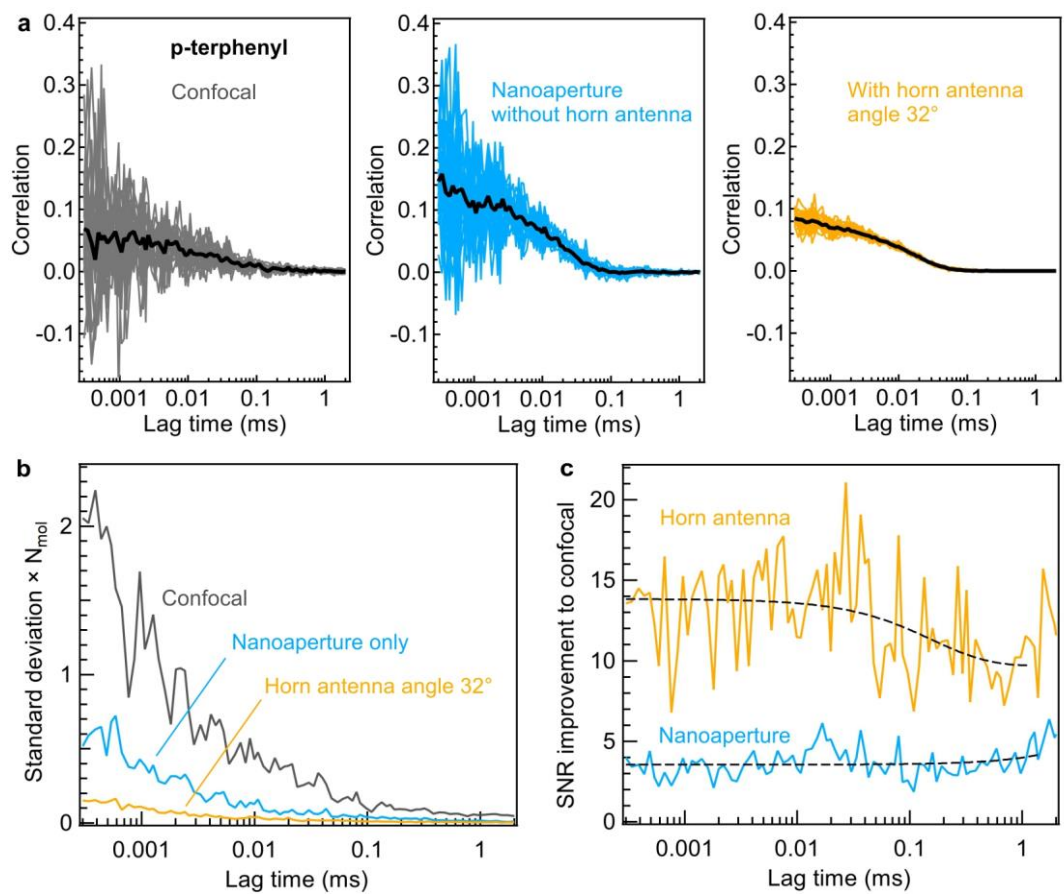

**Figure S7.** High fluorescence brightness per molecules improves the signal to noise ratio in FCS. (a) Raw FCS correlation functions of p-terphenyl for the confocal reference, the single 65 nm nanoaperture and the horn antenna. Each graph shows a superposition of 20 individual FCS curves (thin lines) recorded with 1 s integration time. The spread of these 1 s FCS traces represents the statistical noise of the experimental data. The thick black trace is the average data with 20 s integration time. The higher brightness obtained with the horn antenna directly translates into a reduced noise around the average value without any post-treatment analysis.[4] The concentration for the confocal data shown here is reduced by 1000x. (b) Standard deviation of the FCS trace deduced from the spread of the experimental data points in (a): for each lag time, the standard deviation is calculated among the set of FCS curves recorded with 1 s integration time. The standard deviation is then normalized by the number of molecules $N_{mol}$ deduced from the FCS fit to obtain a concentration-independent quantity representing the noise in an FCS acquisition. These data quantify the noise reduction seen in (a) and the different cases can be directly compared. (c) The normalized standard deviation in (b) is used to compute the signal to noise ratio (SNR) improvement as compared to the confocal case. The noise in FCS depends linearly on the fluorescence brightness per molecule.[4] As this approach is only based on statistical analysis and no numerical fit is performed, this dataset independently confirms the fluorescence enhancement of the brightness per molecule deduced using FCS fitting in Fig. 1g.

## S6. Additional autofluorescence decay traces

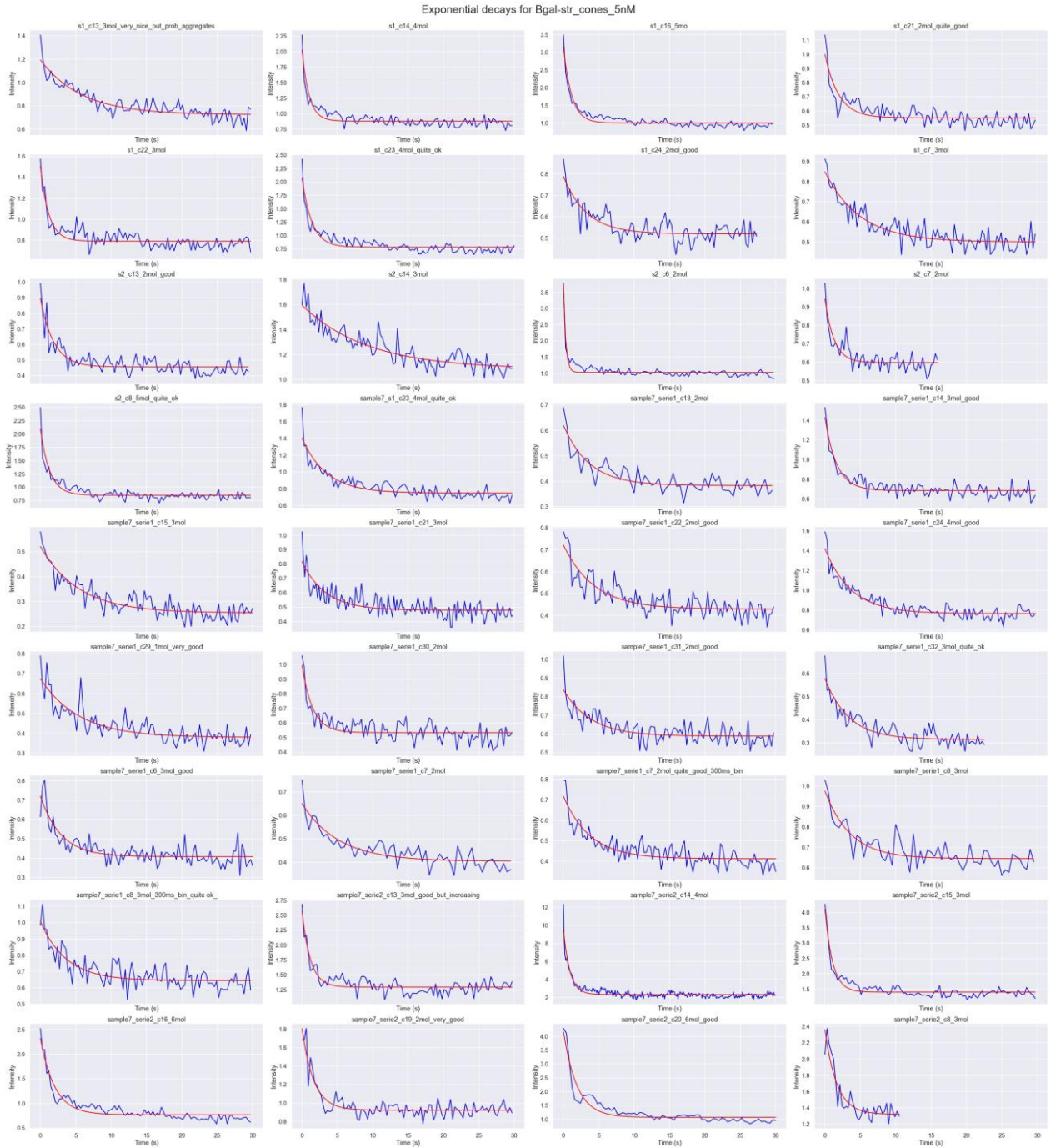

**Figure S8.** Autofluorescence time traces recorded on different individual horn antennas using 5 nM label-free β-galactosidase-streptavidin. The red lines are exponential fits used to extract the decay amplitude. The vertical axis is in kcounts per second.

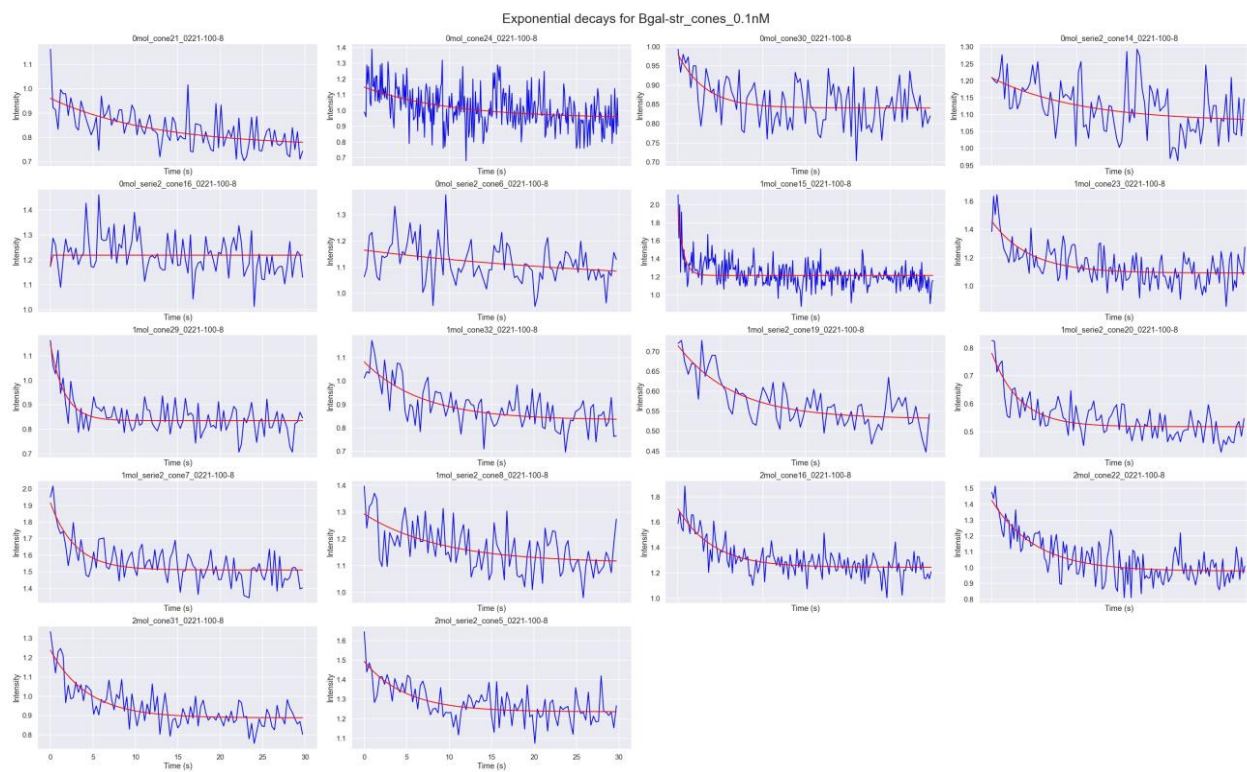

**Figure S9.** Autofluorescence time traces recorded on different individual horn antennas using 0.1 nM label-free β-galactosidase-streptavidin. The red lines are exponential fits used to extract the decay amplitude. The vertical axis is in kcounts per second.

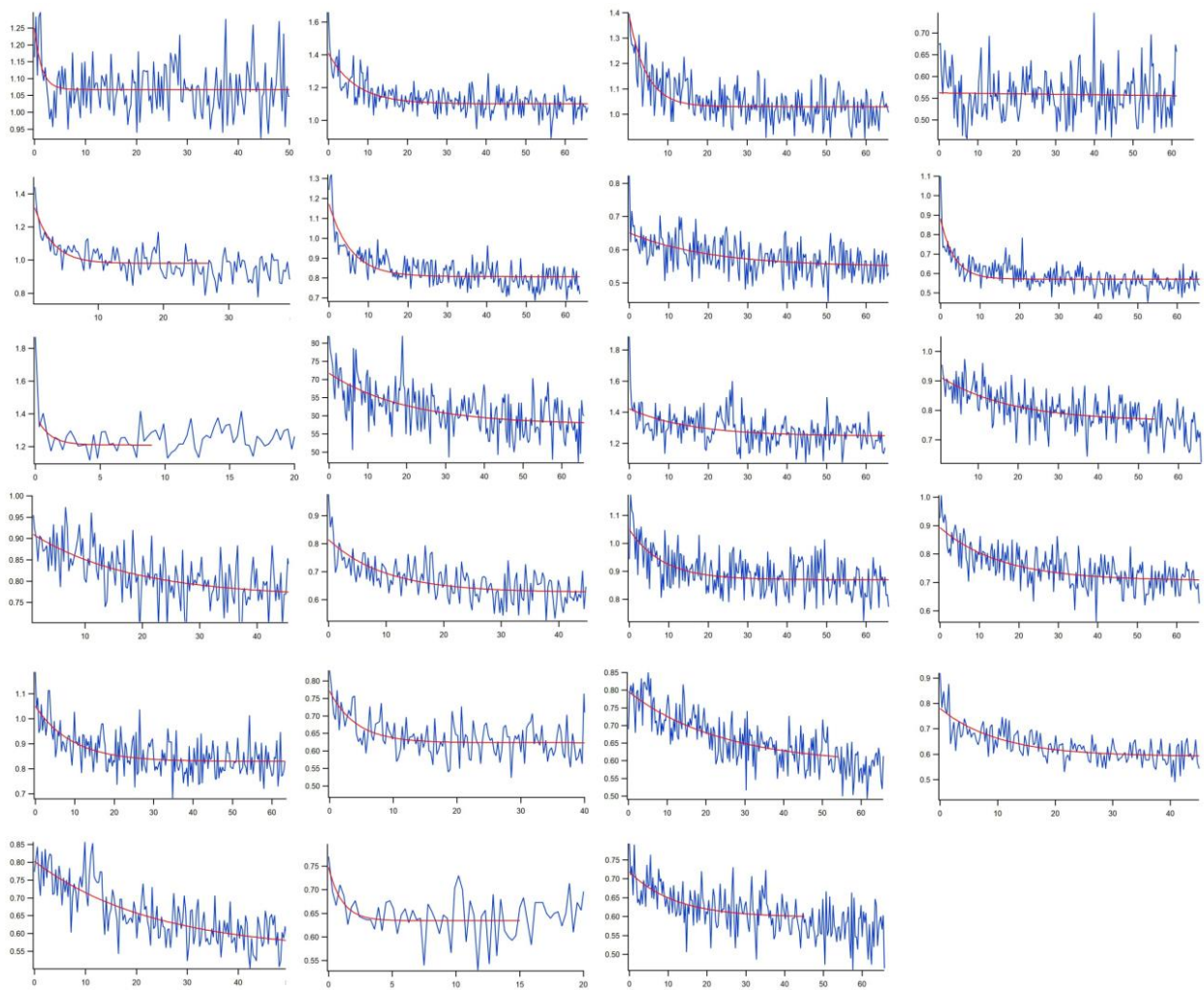

**Figure S10.** Autofluorescence time traces recorded on different individual horn antennas using 5 nM label-free streptavidin. The red lines are exponential fits used to extract the decay amplitude. The vertical axis is in kcounts per second, the horizontal axis is in second.

## S7. Background when no protein is present

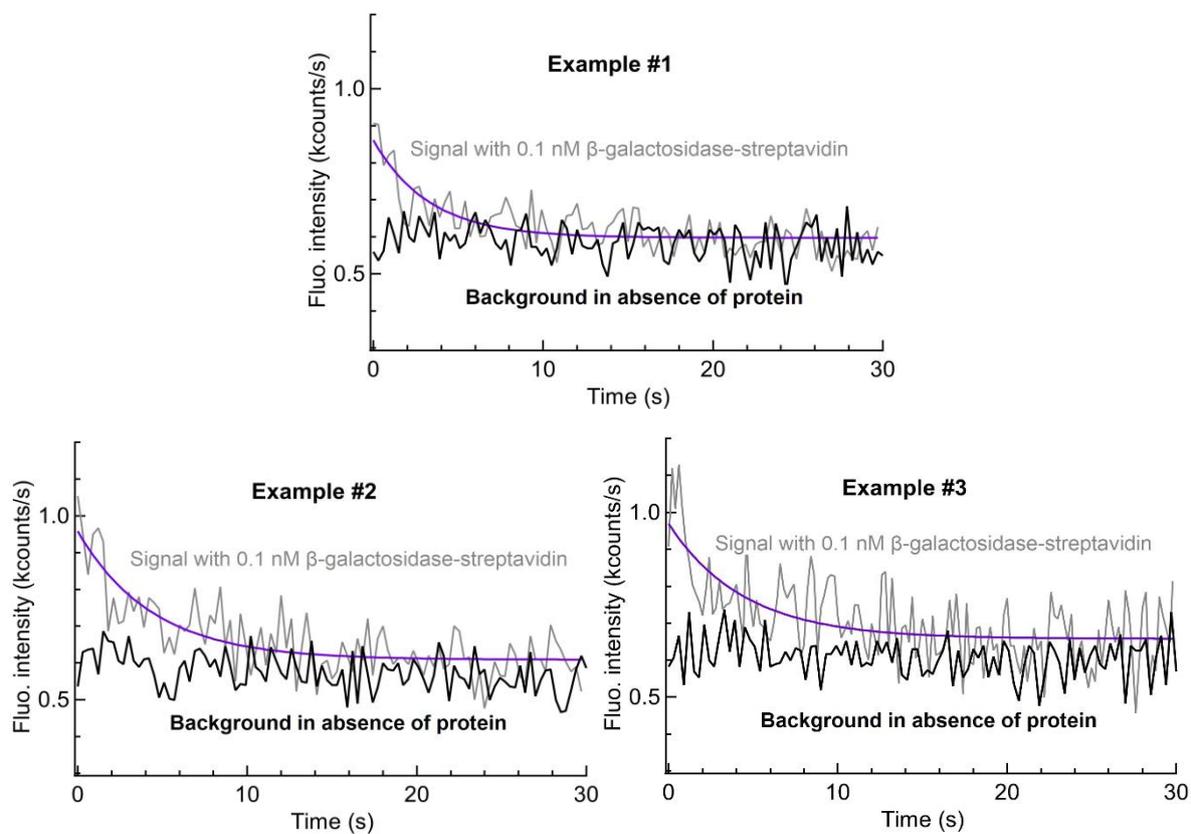

**Figure S11.** Autofluorescence background time traces recorded on different individual horn antennas using only the Hepes buffer solution (including the GODCAT oxygen scavenger system and 10mM DABCO). Typical time traces obtained using 0.1 nM label-free β-galactosidase-streptavidin are displayed for comparison.

## S8. Comparison of single protein autofluorescence time traces with and without horn antenna

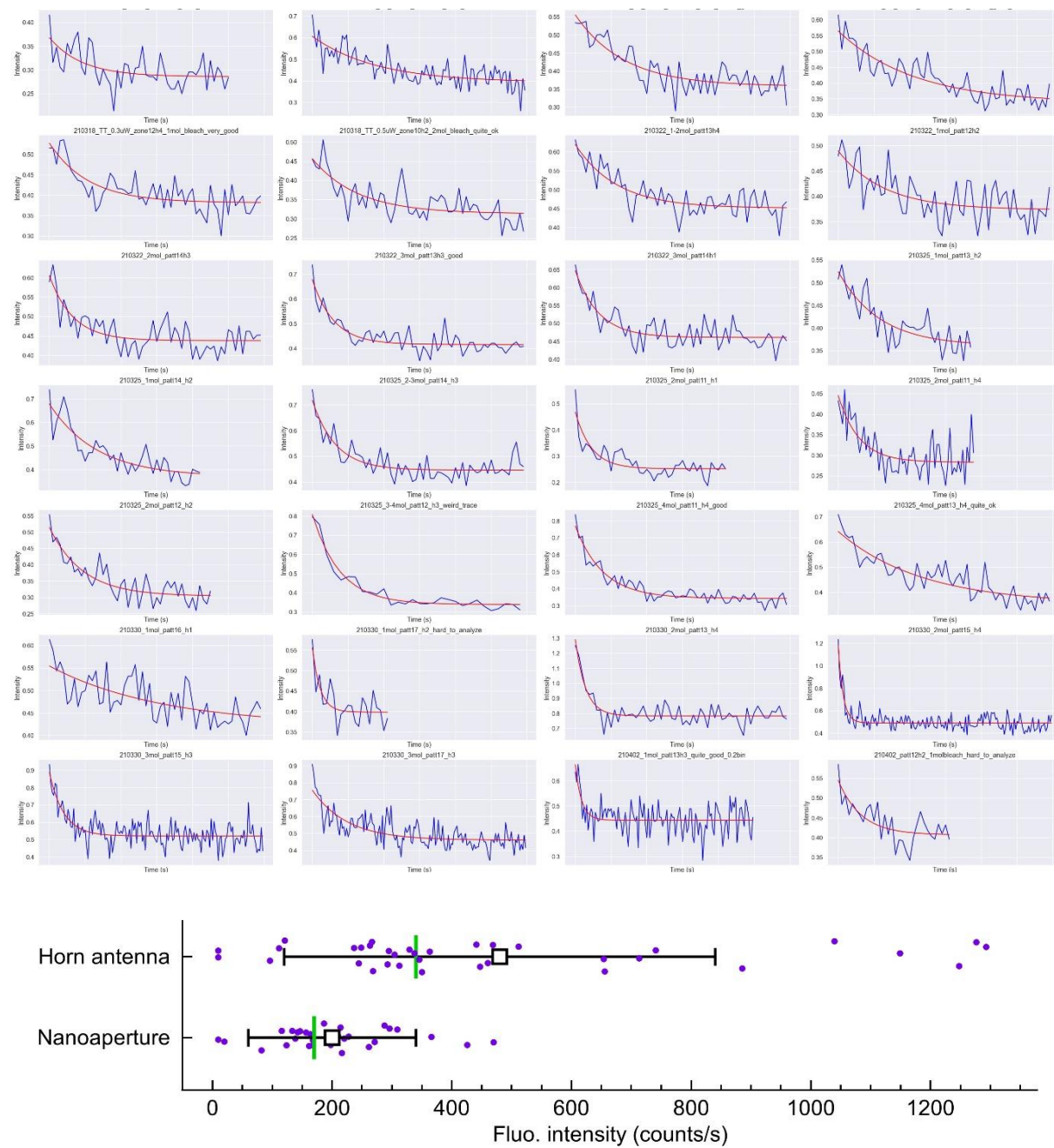

**Figure S12.** Autofluorescence time traces recorded on different individual nanoapertures (without conical reflector) using 5 nM label-free β-galactosidase-streptavidin. The red lines are exponential fits used to extract the decay amplitude. The vertical axis is in kcounts per second. The bottom graph compares the exponential fit amplitudes obtained from individual autofluorescence time traces for the horn antenna and the nanoaperture. The points are vertically shifted using a uniform statistical distribution. The white square marker denotes the average value with the bars extending to one standard deviation. The green vertical line indicates the median. The number of measurements are 35 with the horn antenna and 28 with the nanoaperture. These data confirm the 2 to 3× signal improvement brought by the horn antenna.

## S9. Dependence with the aperture diameter

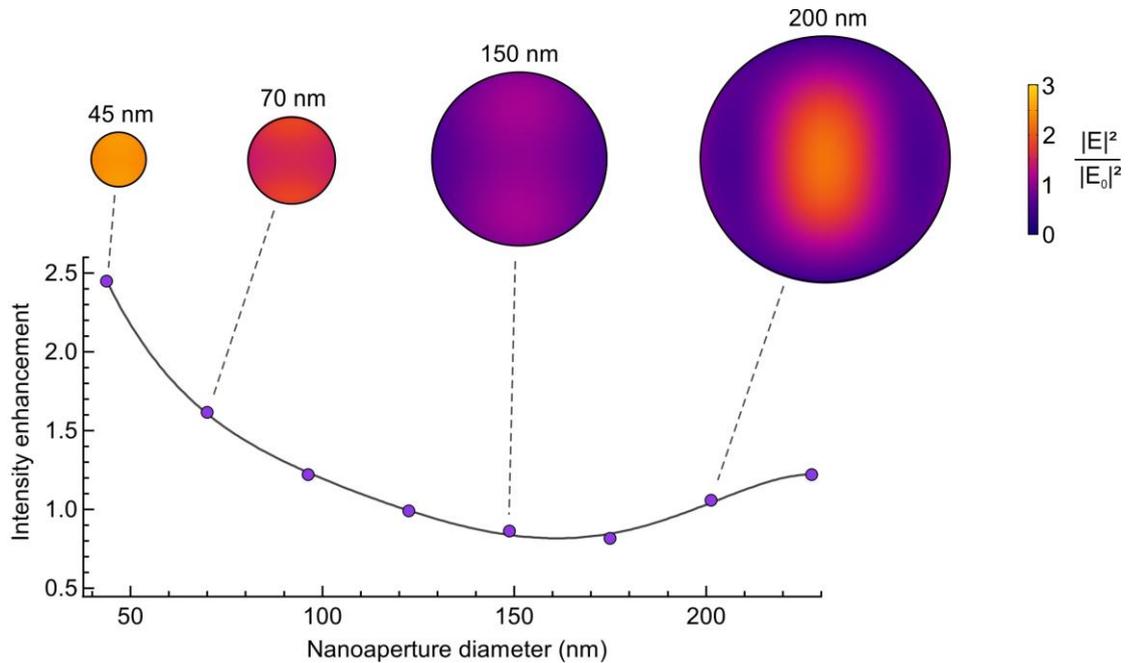

**Figure S13.** Numerical simulations of the 295 nm excitation intensity distributions as a function of the nanoaperture diameter. The plane of interest is taken 50 nm below the quartz-aluminum interface to reproduce the conditions used to detect proteins immobilized at the surface of the 50 nm undercut into the quartz substrate [6,7] (Fig. 2). For clarity only the aperture surface accessible to the proteins is shown here, we do not plot the field distribution into the substrate. The intensity enhancement respective to the incoming intensity is averaged over the whole aperture surface and plotted as a function of the aperture diameter. While smaller diameters feature higher local excitation enhancements, they also induce stronger quenching losses,[8,9] so the optimum diameter to maximize the net signal enhancement is shifted towards slightly larger diameters.[10] The 65 and 200 nm diameters chosen here are based on an experimental screen of the diameter influence and optimization of the signal-to-noise ratio. The grey line is a polynomial interpolation between the simulated data points.

## S10. Proof-of-principle fluorescence lifetime measurements from a single label-free protein

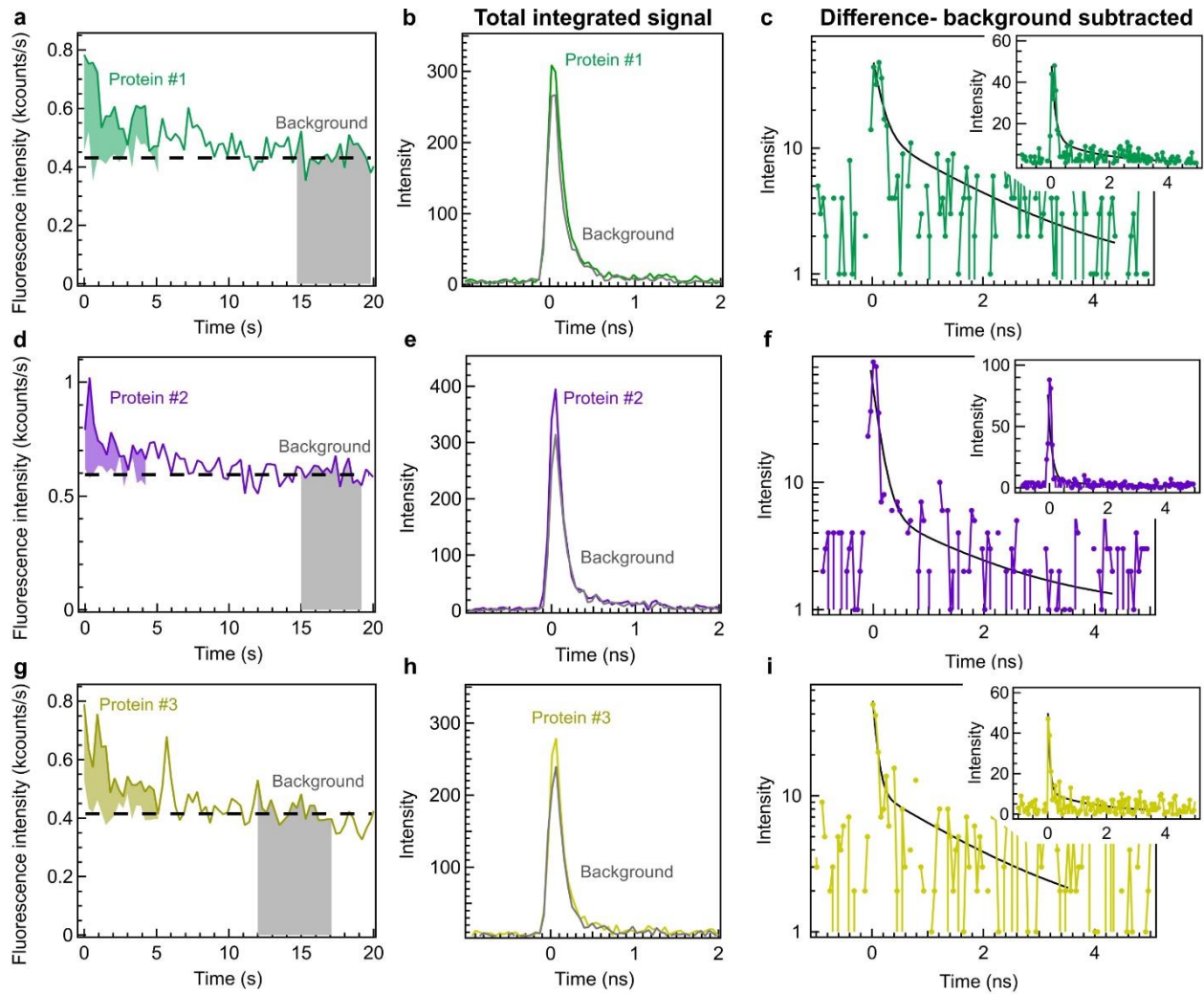

**Figure S14.** Single photon counting histograms from a single protein. We use here the data collected for immobilized β-galactosidase-streptavidin with 0.1 nM concentration. The traces are selected so that the decay amplitude is around 200 to 300 counts per second, which is the most representative of a single protein bleaching (Fig. 2). (a,d,g) Fluorescence time traces used for the analysis. For each selected time interval (shaded areas), the photon arrival times are sorted to compute the fluorescence lifetime decay histograms (b,e,h). The background influence is taken into account by subtracting the histogram taken once the protein of interest has bleached using the same integration time as the protein emission duration (grey traces in a,d,g). After background subtraction (c,f,i), the histogram contains the information representative of the UV photons emitted by a single protein. The histogram binning time is 48 ps. The insets in (c,fi) show the histograms on a linear vertical scale. The fit results are detailed in Tab. S3.

Measuring the fluorescence lifetime from a single molecule remains challenging, even with bright fluorescent dyes in the visible, as the limited photon budget before photobleaching is split into the different histogram channels. The data shown here demonstrates that despite the technical challenges, the high collection efficiency brought by the UV horn antenna enables extracting the lifetime histogram from a single label-free protein.

**Table S3.** Fit parameters for the single molecule fluorescence lifetime data in Fig. S14c,f,i. Here the 10 ps background contribution has been removed by the subtraction treatment. However, the data acquired while averaging at high µM concentration (Fig. S15b) shows that the lifetime decays of β-galactosidase are not a single exponential, but rather a bi-exponential with a long and a short component (the decays for β-galactosidase and β-galactosidase-streptavidin are similar). This multi-exponential response is typical of protein autofluorescence.[11] Owing to the limited total photon budget from a single protein, the statistical noise is large for the component with the longest lifetime. In the lifetime analysis, we have decided to fix the long lifetime component to its 1.6 ns value obtained from the averaging at 1.7 µM (Fig. S15b, Tab. S5). The lifetimes are expressed in ns, the intensities are normalized so that their sum equals 1. $\tau^{av}_{1-2}$ denotes the intensity-averaged lifetime. The instrument response function is 120 ps (full width at half-maximum) taking into account the 48 ps binning time for the lifetime histogram.

|  | $\tau_1$ | $\tau_2$ (fixed) | $I_1$ | $I_2$ | $\tau^{av}_{1-2}$ |
|---|---|---|---|---|---|
| Protein #1 | 0.14 | 1.6 | 0.78 | 0.22 | 0.5 ± 0.3 |
| Protein #2 | 0.13 | 1.6 | 0.90 | 0.10 | 0.3 ± 0.2 |
| Protein #3 | 0.07 | 1.6 | 0.83 | 0.17 | 0.3 ± 0.2 |

## S11. Fluorescence enhancement of diffusing β-galactosidase proteins with horn antennas

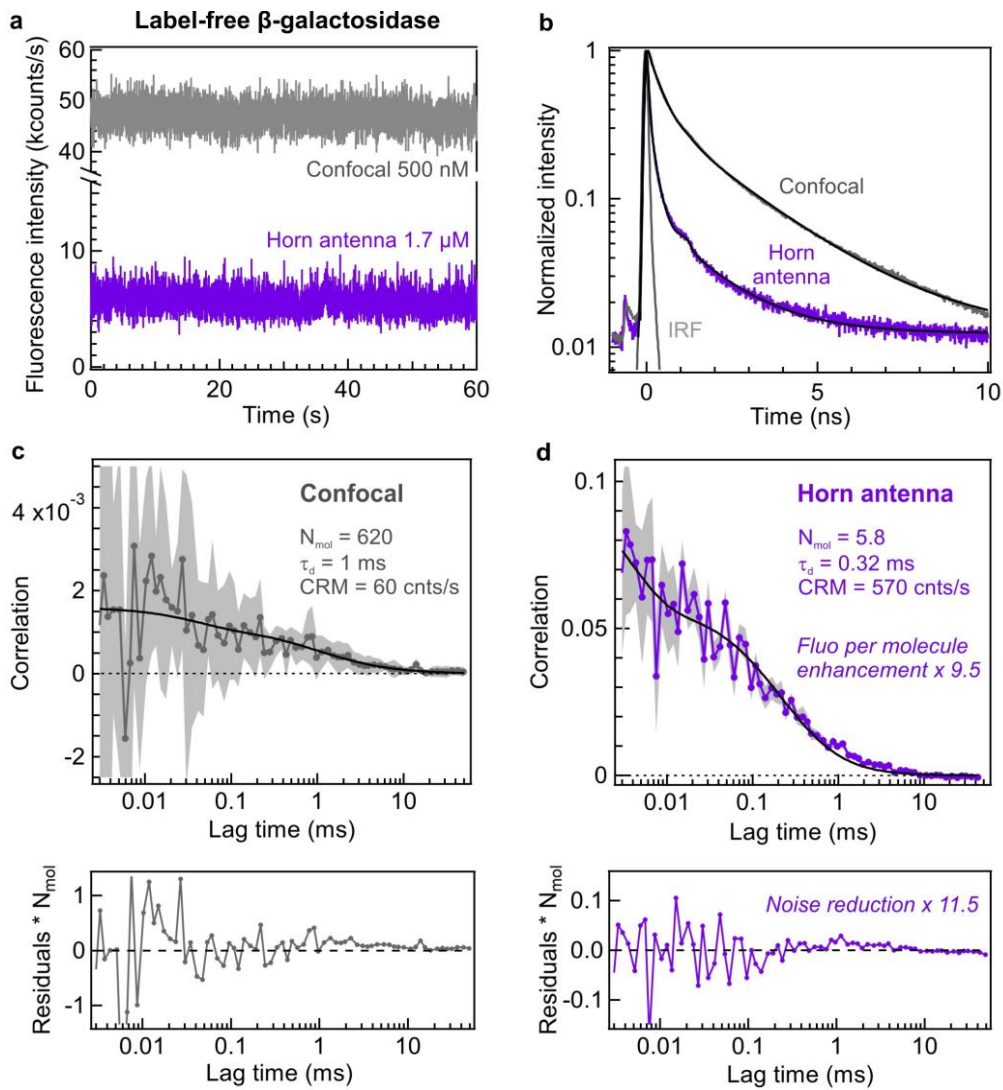

**Figure S15.** Demonstration of fluorescence enhancement for label-free β-galactosidase proteins inside a horn antenna with 32° cone angle and 65 nm aperture diameter. (a) Fluorescence time trace at 10 µW laser power with 10 ms bin time. The background for the confocal is 8 kHz and 2.4 kHz for the horn antenna. The background for the confocal experiment stems mostly from the UV fluorescence of the GODCAT oxygen scavenger in the 2 fL confocal volume. For the horn antenna, the GODCAT background is reduced as the detection volume is ~400× smaller, but an additional contribution comes from the photoluminescence of the metal. (b) Fluorescence lifetime decay and numerical fits. IRF is our instrument response function (full width at half maximum 160 ps). The fit results are summarized in Tab. S5. (c,d) FCS correlation functions (dots) and their numerical fits (black curves) corresponding to the time traces in (a). The grey shaded traces indicate the noise of the FCS data. The FCS fit results are summarized in Tab. S4 and the main quantities are indicated on the graph. The lower traces show the residuals from the fit functions multiplied by the number of detected molecules (inverse of correlation amplitude). These normalized residuals do not depend on the protein concentration so that their amplitudes can be directly compared. The acquisition times are identical here with 130 s. To quantify the reduction of the normalized residuals, we sum their absolute values in the 10 µs - 1 ms lag time

interval. We observe a 9.5 ± 2.0 × increase of the CRM brightness per molecule. This fluorescence enhancement is supported by the 11.5× reduction of the normalized residuals amplitude as the noise in FCS scales directly proportional to the brightness per emitter.[4] As a side note, our data indicates that a fast blinking contribution ($n_T$, $\tau_T$) is needed to yield flat residuals. The precision is limited for these two parameters, but they do not have any influence on our main conclusions. Different processes could explain this contribution in the sub-10 µs range such as dark state blinking, fast protein conformation changes, metal quenching in close nanometer proximity to the surface, and/or residual contribution from the photodetector afterpulsing. Currently we cannot distinguish between these potential sources, but the photokinetics in the sub-10 µs range do not impact any of our main conclusions.

**Table S4.** Fit parameters for the FCS data displayed on Fig. S15c,d. The shape parameter κ is fixed at 8 for the confocal case and 1 for the aperture and horn antennas. The 295 nm laser power is 10 µW. The β-galactosidase concentration is 500 nM for the confocal case and 1.7 µM for the horn antenna.

|  | $F$ (kHz) | $B$ (kHz) | $N_{mol}$ | $n_T$ | $\tau_T$ (µs) | $\tau_d$ (ms) | CRM (Hz) | Fluo. enhancement |
|---|---|---|---|---|---|---|---|---|
| Confocal | 47.1 | 8 | 620 | 0.4 | 40 | 1 | 60 ± 10 | - |
| Horn antenna 32° | 5.7 | 2.4 | 5.8 | 0.75 | 4 | 0.32 | 570 ± 30 | 9.5 ± 2.0 |

**Table S5.** Fit parameters for the fluorescence lifetime data in Fig. S15b. The lifetimes are expressed in ns, the intensities are normalized so that their sum equals 1. $\tau_{2-3}^{av}$ denotes the intensity-averaged lifetime of the 2$^{nd}$ and 3$^{rd}$ components (discarding the 10 ps background contribution) and is used as a read-out lifetime to demonstrate the lifetime reduction with the optical horn antenna.

|  | $\tau_1$ | $\tau_2$ | $\tau_3$ | $I_1$ | $I_2$ | $I_3$ | $\tau_{2-3}^{av}$ |
|---|---|---|---|---|---|---|---|
| Confocal | 0.01 | 0.39 | 2.38 | 0.07 | 0.24 | 0.69 | 1.9 ± 0.1 |
| Horn antenna 32° | 0.01 | 0.20 | 1.61 | 0.48 | 0.20 | 0.32 | 1.1 ± 0.1 |

## S12. β-galactosidase autofluorescence spectra in presence of urea

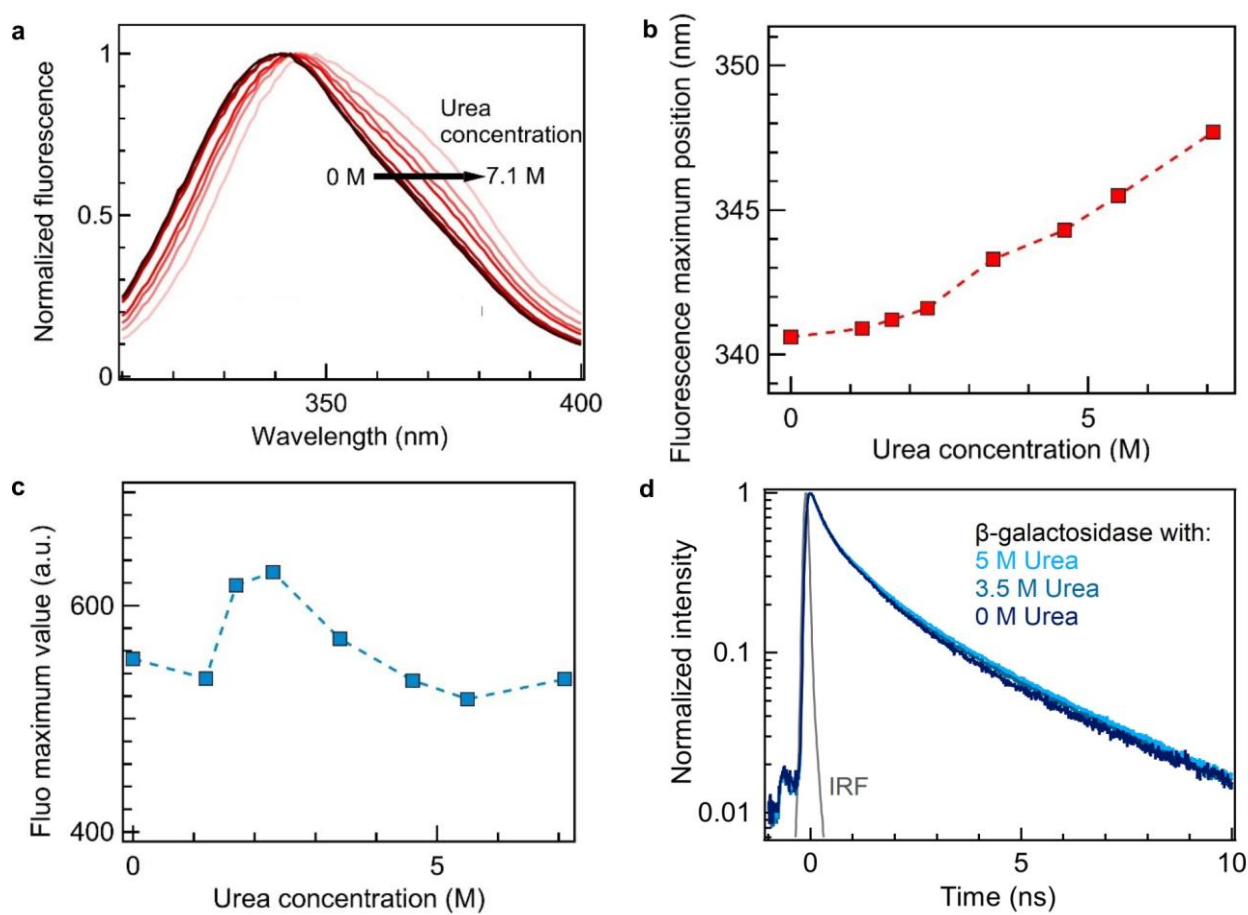

**Figure S16.** (a) Autofluorescence spectra of β-galactosidase for increasing concentrations of urea. Upon unfolding and denaturation, the autofluorescence spectra maximum shifts towards larger wavelengths (b), while the maximum intensity at the fluorescence peak stays rather similar (c). These spectra were recorded on a Tecan Spark 10M spectrofluorometer with excitation fixed at 260 nm. The β-galactosidase concentration is 700 nM in a 300 mM NaCl, 25 mM Hepes, 0.5 % Tween20, pH 7 buffer solution. (d) Normalized fluorescence lifetime decay traces for β-galactosidase with increasing urea concentrations. No major change is detected on the fluorescence decay, the fluorescence lifetime increases by a few percent in presence of 5 M urea.

## S13. Diffusion time of β-galactosidase and viscosity calibration in presence of urea

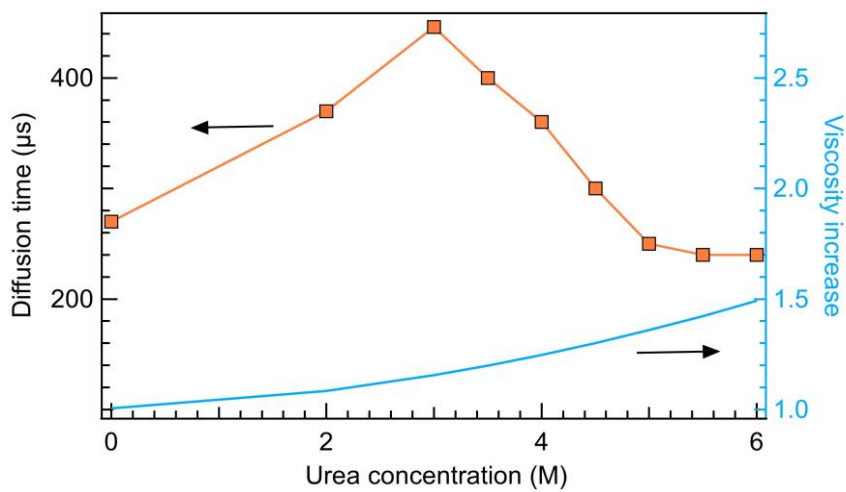

**Figure S17.** FCS diffusion time of β-galactosidase recorded on the horn antenna in presence of urea (left axis). This data is used to compute the hydrodynamic radius of the protein upon unfolding and denaturation shown in Fig. 4b following the approach described in [10]. As the hydrodynamic radius also depends on the viscosity of the solution, we have calibrated the viscosity increase due to the addition of urea (blue trace, right axis). This calibration is performed separately using FCS on a visible-light confocal microscope with Alexa 647 fluorescent molecules.[3]

**Supplementary references**

1. Balanis, C. A. *Antenna theory: analysis and design*. (John Wiley & Sons, 2005).
2. Novotny, L. & Hulst, N. van. Antennas for light. *Nature Photonics* **5**, 83–90 (2011).
3. Baibakov, M. *et al.* Extending Single-Molecule Förster Resonance Energy Transfer (FRET) Range beyond 10 Nanometers in Zero-Mode Waveguides. *ACS Nano* **13**, 8469–8480 (2019).
4. Wenger, J. *et al.* Nanoaperture-Enhanced Signal-to-Noise Ratio in Fluorescence Correlation Spectroscopy. *Anal. Chem.* **81**, 834–839 (2009).
5. Loeff, L., Kerssemakers, J. W. J., Joo, C. & Dekker, C. AutoStepfinder: A fast and automated step detection method for single-molecule analysis. *Patterns* **2**, 100256 (2021).
6. Tanii, T. *et al.* Improving zero-mode waveguide structure for enhancing signal-to-noise ratio of real-time single-molecule fluorescence imaging: A computational study. *Phys. Rev. E* **88**, 012727 (2013).
7. Wu, M. *et al.* Fluorescence enhancement in an over-etched gold zero-mode waveguide. *Opt. Express* **27**, 19002–19018 (2019).
8. Jiao, X., Peterson, E. M., Harris, J. M. & Blair, S. UV Fluorescence Lifetime Modification by Aluminum Nanoapertures. *ACS Photonics* **1**, 1270–1277 (2014).
9. Jiao, X., Wang, Y. & Blair, S. UV fluorescence enhancement by Al and Mg nanoapertures. *J. Phys. D: Appl. Phys.* **48**, 184007 (2015).
10. Barulin, A., Claude, J.-B., Patra, S., Bonod, N. & Wenger, J. Deep Ultraviolet Plasmonic Enhancement of Single Protein Autofluorescence in Zero-Mode Waveguides. *Nano Lett.* **19**, 7434–7442 (2019).
11. Lakowicz, J. R. *Principles of Fluorescence Spectroscopy*. (Springer US, 2006).